\documentclass[%
 reprint,
 amsmath,amssymb,
 aps,
 pre,
 showkeys
]{revtex4-1}

\usepackage{lipsum,amsmath}
\usepackage{mathtools}
\usepackage[]{graphics}
\usepackage{graphicx}
\usepackage{dcolumn}

\usepackage{tikz}
\usepackage{tikz-3dplot}
\usetikzlibrary{shapes, fadings}
\usetikzlibrary{calc}
\usetikzlibrary{patterns}
\usetikzlibrary{3d}

\usetikzlibrary{arrows, decorations.markings}

\makeatletter
\tikzoption{canvas is plane}[]{\@setOxy#1}
\def\@setOxy O(#1,#2,#3)x(#4,#5,#6)y(#7,#8,#9)%
  {\def\tikz@plane@origin{\pgfpointxyz{#1}{#2}{#3}}%
   \def\tikz@plane@x{\pgfpointxyz{#4}{#5}{#6}}%
   \def\tikz@plane@y{\pgfpointxyz{#7}{#8}{#9}}%
   \tikz@canvas@is@plane
  }
\makeatother

\newcommand{\bnabla}{\mbox{\boldmath$\nabla$}}

\tikzfading[name=fade out,
            inner color=transparent!0,
            outer color=transparent!100]

\tikzstyle{vecArrow} = [thick, decoration={markings,mark=at position
   1 with {\arrow[semithick]{open triangle 60}}},
   double distance=1.4pt, shorten >= 5.5pt,
   preaction = {decorate},
   postaction = {draw,line width=1.4pt, white,shorten >= 4.5pt}]
\tikzstyle{innerWhite} = [semithick, white,line width=1.4pt, shorten >= 4.5pt]

\usepackage{csvsimple}

\usetikzlibrary{shapes.misc}

\tikzset{cross/.style={cross out, draw=black, minimum size=2*(#1-\pgflinewidth), inner sep=0pt, outer sep=0pt},
cross/.default={1pt}}

\begin{document}



\title{Dynamo action in sliding plates of anisotropic electrical conductivity}

\author{T.~Alboussi\`ere}
\author{K.~Drif}
\affiliation{Univ. Lyon, CNRS, Universit\'e Claude Bernard  Lyon 1, Ens de Lyon, UMR 5276 LGL-TPE, F-69622 Villeurbanne, France}
\author{F.~Plunian}
\affiliation{Univ. Grenoble Alpes, CNRS, UMR 5275 ISTerre, F-38041 Grenoble, France}

\date{\today}

\begin{abstract}
With materials of anisotropic electrical conductivity, it is possible to generate a dynamo with a simple velocity field, of the type precluded by Cowling's theorems with isotropic materials. Following a previous study by Ruderman and Ruzmaikin 
\cite{RR84} who considered the dynamo effect induced by a uniform shear flow, we determine the conditions for the dynamo threshold when a solid plate is sliding over another one, both with anisotropic electrical conductivity. We obtain numerical solutions for a general class of anisotropy and obtain the conditions for the lowest magnetic Reynolds number, using a collocation Chebyshev method. 
	In a particular geometry of anisotropy and wavenumber, we also derive an analytical solution, where the eigenvectors are just combinations of four exponential functions. An explicit analytical expression is obtained for the critical magnetic Reynold number. Above the critical magnetic Reynold number, we have also derived an analytical expression for the growth rate showing that this is a 'very fast' dynamo, extrapolating on the 'slow' and 'fast' terminology introduced by Vainshtein and Zeldovich \cite{VZ72}.  
\end{abstract}


\maketitle


\section{Introduction}
\label{intro}

Dynamo action is now widely accepted as a mechanism capable of generating the magnetic field of natural objects such as the Earth, other planets, the Sun, all stars, the solar wind,  the interstellar medium... Since Herzenberg \cite{herzenberg} and Backus \cite{backus}, we have example of mathematical dynamos, however those solutions are not very easy to describe and to teach. The dynamo proposed by Ponomarenko \cite{ponomarenko} is perhaps the simplest case. Any simple dynamo configuration, easy to derive analytically and with an easy mechanism to grasp is welcome.

Since the pionniering paper of Cowling \cite{cowling33}, we know that a magnetic field generated by dynamo action cannot be too simple. We also know that a velocity field with too many symmetries cannot sustain dynamo action. For instance, a planar flow is found to be unable to maintain a dynamo \cite{zeldovich57,ZR80}. However, these conclusions are always associated with a material of isotropic electrical conductivity. For instance, Ruderman and Ruzmaikin \cite{RR84} consider a planar shear flow (uniform shear) and a simple anisotropic tensor of electrical conductivity, with one direction having a different value than the other two. They obtain dynamo action with such a simple shear flow, provided the direction of conductivity anisotropy is not aligned with the flow direction nor with the direction of the gradient of the flow. Using an asymptotic approximation, they show that the configuration is a fast dynamo, {\it i.e.} the growth rate does not vanish as the magnetic Reynolds number is increased toward infinity.  

In the present work, we also consider a shear flow and a similar conductivity tensor as in \cite{RR84}. Our motivation is to obtain the simplest possible configuration, with the simplest possible analytical derivation of the critical magnetic Reynolds number. In this respect, we found that the best case is to have a localized shear (Dirac function) between two 'plates' of uniform velocity, sliding on top of each other. In each plate, the induction equation -- with anisotropic conductivity -- leads to elementary solutions. The global dynamo solution is then obtained by applying boundary conditions, including continuity conditions at the interface between the plates.

\section{Configuration}
\label{config}

\begin{figure}
\begin{center}
	\includegraphics[width=8.5cm,keepaspectratio]{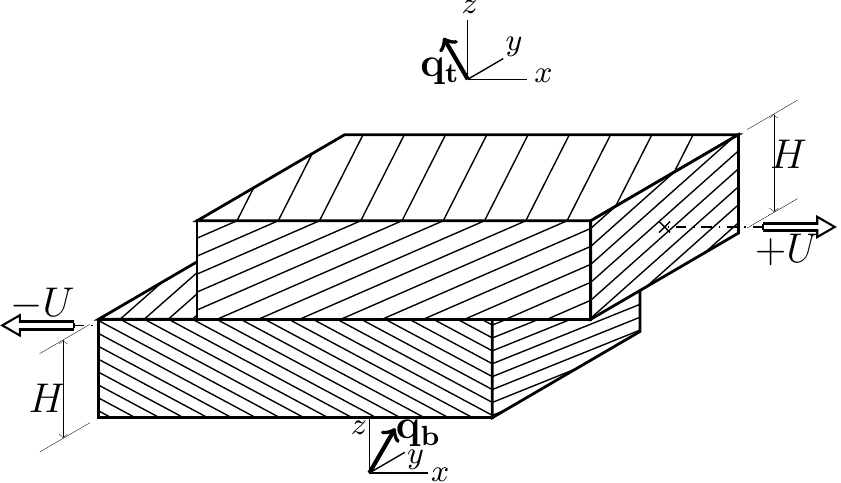}
\caption{Two plates of finite and equal thickness $H$ slide on each other with relative velocity $\pm U$. Each plate has a uniform anisotropic electrical conductivity, characterized by a lower value in the direction ${\bf q}$. The thin lines drawn on each cut-section of the plates correspond to the directions of large electrical conductivity. }
\label{schema}
\end{center}
\end{figure}

We consider two plates of thickness $H$, put on top of each other, and sliding relative to each other with a velocity $\pm U$ (see Fig.~\ref{schema}). A frame of reference $(x,y,z)$ is defined with $x$ along the sliding direction, $z$ along the direction perpendicular to the plates and $y$ completes the direct orthogonal Cartesian frame. The origin $z=0$ is taken at the interface between the plates. The plates have an anisotropic electrical conductivity: one direction, denoted by the unitary vector ${\bf q}$, has a lower conductivity $\sigma _1$, while the other two perpendicular principal directions of the conductivity tensor have a large electrical conductivity $\sigma _0$.

\begin{figure}
\begin{center}
	\includegraphics[keepaspectratio]{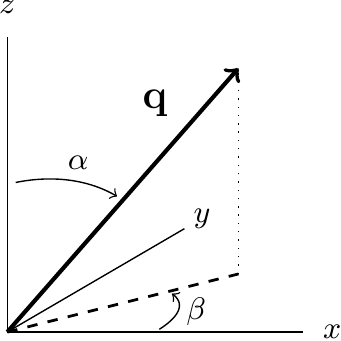}
\caption{Angles $\alpha$ and $\beta$ define the orientation of the anisotropy. The angle between the $z$ axis and ${\bf q}$ is denoted $\alpha$ while $\beta$ denotes the angle between the $x$ axis and the projection of ${\bf q}$ in the $(x,y)$ plane: $q_x = \cos \beta \sin \alpha$, $q_y = \sin \beta \sin \alpha$ and $q_z = \cos \alpha$.}
\label{angles}
\end{center}
\end{figure}

The unitary vector ${\bf q}$ relative to the direction of anisotropy is itself defined through two angles, $\alpha$ and $\beta$ (see Fig.~\ref{angles}). One, $\alpha$, is the angle between the $z$ axis and the vector ${\bf q}$ while the other, $\beta$, is the angle between the $x$ axis and the projection of ${\bf q}$ onto the $(x,y)$ plane 
\begin{align}
q_x &= \cos \beta \sin \alpha , \label{qx} \\
q_y &= \sin \beta \sin \alpha , \label{qy} \\
q_z &= \cos \alpha . \label{qz} 
\end{align}
The tensor of electrical conductivity $\Sigma$ takes the following form, which is the general form of a positive definite tensor with two equal eigenvalues (see \cite{RR84})
\begin{equation}
\Sigma_{ij} = \sigma _0 \delta _{ij} + \left( \sigma _1 - \sigma _0 \right) q_i q_j . \label{tensorcond}
\end{equation}
Its inverse, the electrical resistivity tensor $R$, will actually be more useful and its expression is the following
\begin{equation}
R_{ij} = \frac{1}{\sigma _0} \delta _{ij} + \left( \frac{1}{\sigma _1} - \frac{1}{\sigma _0} \right) q_i q_
j . \label{tensorresist}
\end{equation}
The only requirement, from the second law of thermodynamics, is that both $\sigma _0$ and $\sigma _1$ have positive values. 

\section{Governing equations}
\label{gov}

The kinematic dynamo problem is entirely expressed in the induction equation, governing the evolution of the divergence-free magnetic field ${\bf B}$
\begin{equation}
\frac{\partial {\bf B} }{\partial t} = {\bf \bnabla} \times \left( {\bf u} \times {\bf B} \right) - {\bf \bnabla} \times \left( \eta \cdot {\bf \bnabla} \times {\bf B} \right) , \label{induction}
\end{equation}
where $\eta = (\mu \Sigma )^{-1}$ is the anisotropic tensor of magnetic diffusivity and $\mu$ is the magnetic permeability. We consider materials without any particular magnetic properties and the magnetic permeability is that of vacuum $\mu _0 = 4 \pi \, 10^{-7}$~H~m$^{-1}$.  

We use the dimensional scales of $H$, $\left( \mu _0 \sigma _0 \right) ^{-1}$, $H^2 \mu _0 \sigma _0$ and $\left( \mu _0 \sigma _0 H \right) ^{-1}$ for distance, magnetic diffusivity, time and velocity respectively, so that the dimensionless induction equation takes the exact same expression as equation (\ref{induction}), except that the dimensionless diffusivity tensor $\eta$ is now
\begin{equation}
\eta _{ij} = \delta _{ij} + \eta _1 q_i q_j , \label{diffusivity}
\end{equation}
where $\eta _1 = \sigma _0 / \sigma _1 -1$ must be larger than $-1$, since $\sigma _0$ and $\sigma _1$ are only required to be positive. The dimensionless value of the imposed velocity will correspond to the dimensionless magnetic Reynolds number, $Rm$. 

The divergence-free magnetic induction ${\bf B}$ is expressed using poloidal and toroidal scalars $P$ and $T$ 
\begin{equation} \label{PoloidalToroidal}
{\bf B} = {\bf \bnabla} \times \left( T {\bf e}_z \right) + {\bf \bnabla} \times \left( {\bf \bnabla} \times \left( P {\bf e}_z \right) \right), 
\end{equation}
where $P$ and $T$ are the poloidal and toroidal scalar functions. 
Because of the invariance of the problem in the $x$ and $y$ directions and in time $t$, we look for dynamo solutions as a series of eigenvectors of the following form
\begin{align} 
P &=  P(z) \exp \left(i k_x x + i k_y y + \gamma t \right), \label{Peig} \\ 
T &=  T(z) \exp \left(i k_x x + i k_y y + \gamma t \right), \label{Teig}
\end{align}
where $k_x$ and $k_y$ are the (real) wavenumbers in the $x$ and $y$ directions and $\gamma$ is the growth rate of the mode. A dynamo mode is obtained when the real part of $\gamma$ is positive. We re-use the same symbols $P$ and $T$ for the $z$-dependent functions entering the expressions for the poloidal and toroidal parts.  

In terms of methods, we solve the induction equation in each plate, and then consider the boundary conditions applying to the solutions, including the relative sliding condition between the plates. So, in a first step, we seek to solve the induction equation in a domain of uniform dimensionless velocity $U$ and uniform anisotropy defined from the angles $\alpha$ and $\beta$ and from the anisotropic factor $\eta_1$ (see equation (\ref{diffusivity})). The uniform velocity is denoted generically $U$, but in effect it will be $U$ for the upper plate and $-U$ for the lower plate. 
The induction equation and its curl in the $z$ direction provide two coupled equations for $P$ and $T$ (see appendix \ref{derivationPT})
\begin{align}
\begin{split}\label{eqP}
\gamma P = {} & -i k_x U P + \left( 1 + \eta _1 \frac{k_q^2}{k^2} \right) \left[ P'' - k^2 P \right] \\ 
&  - i \eta _1 k_q q_z T + \eta _1  \frac{k_q d_q}{k^2} T' , 
\end{split} \\
\begin{split}\label{eqT}
\gamma T = {} & - i k_x U T + T'' - k^2 \left( 1 + \eta _1 q_z^2 \right) T \\ 
& - 2 i \eta _1 d_q q_z T' + \eta _1 \frac{d_q^2}{k^2} T'' \\
& -  i \eta _1 q_z k_q \left[ P'' - k^2 P \right] + \eta _1 \frac{k_q d_q}{k^2} \left[ P''' - k^2 P' \right] , 
\end{split} 
\end{align}
where we have introduced the following notations
\begin{align}
k^2 &= k_x^2+k_y^2 , \label{k2} \\
k_q &= k_x q_y -k_y q_x, \label{kq} \\
d_q &= k_x q_x + k_y q_y. \label{divq} 
\end{align}
The $T$ equation (\ref{eqT}) contains a third derivative $P'''(z)$, which requires too many boundary conditions to solve. However, it can be seen on equation (\ref{eqP}) that $P''$ can be obtained in terms of $P$, $T$ and $T'$, so that $P'''$ is expressed in terms of $P'$, $T'$ and $T''$, before it is substituted in equation (\ref{eqT}), which now takes the following form
\begin{equation} \label{eqTnew}
\begin{split}
\gamma \left( T -  \delta P' \right) & =   - i k_x U T + T'' - k^2 T  - \eta _1 k^2 q_z^2 T \\
 - & i \eta _1 q_z d_q \frac{ 2 k^2 + \eta _1 k_q^2}{k^2+\eta _1 k_q^2} T' + \eta _1 \frac{d_q^2}{k^2 + \eta _1 k_q^2} T'' \\
 + & i k_x \delta U P' - i \eta _1 q_z k_q \left[ P'' - k^2 P \right] ,
\end{split} 
\end{equation}
with one more notation
\begin{equation}
\delta = \frac{\eta _1 k_q d_q}{k^2 + \eta _1 k_q^2}. \label{delta}
\end{equation}
Equations (\ref{eqP}) and (\ref{eqTnew}) are the basic eigenvalue equations, that have to be satisfied both in the top plate (subscript $_t$) and in the bottom plate (subscript $_b$). In addition, $P_t$, $T_t$, $P_b$ and $T_b$ have to satisfy boundary conditions. 

\section{Boundary conditions}
\label{bc}

With a second-order differential eigenvalue problem involving four fields, $P_t$, $T_t$, $P_b$ and $T_b$, we need a total of eight boundary conditions. They are the following:
\begin{itemize}
\item $T=0$ at the top of the top plate and at the bottom of the bottom plate: two boundary conditions 
\item $P'=\mp k P$ at the top of the top plate and at the bottom of the bottom plate: two boundary
 conditions
\item $T$, $P$ and $P'$ are continuous at the contact between the plates: three boundary conditions
\item The tangential components of the electrical field are continuous at the contact between the plates: we shall see that it corresponds to one boundary condition only (and not two as expected). 
\end{itemize}

The first condition $T=0$ is related to the continuity of the electric current density. Since there is no electric current in the free space (air or vacuum) above the top plate and below the bottom plate, $j_z$ must be zero on these boundaries, hence $T=0$ according to (\ref{jz}):
\begin{align}
{T_t}_{\left| z=1 \right.} &= 0 , \label{clTtop} \\
{T_b}_{\left| z=-1\right.}  &= 0  . \label{clTbot} 
\end{align}

The second condition $P'=\mp k P$ is a classical condition on insulating boundaries. In the semi-infinite spaces above and below the plates, the harmonic equation $P''=k^2 P$ applies. However we restrict the solutions to decay at infinity, otherwise magnetic energy would come from elsewhere and the plates might not necessarily be responsible for dynamo action: this imposes $P'=-kP$ above and $P'=kP$ below. Next, both $P$ and $P'$ are continuous at the top and bottom interfaces, due to the continuity of the normal component $B_z$ (\ref{Bz}) and continuity of the tangential components $B_x$ and $B_y$ (see (\ref{Bx}) and (\ref{By}) with $T$ continuous from the first boundary conditions). Hence, we have 
\begin{align}
{P'_t}_{\left| z=1 \right.} &= -k {P_t}_{\left| z=1 \right.} , \label{clPtop} \\
{P'_b}_{\left| z=-1 \right.}  &= k {P_b}_{\left| z=-1 \right.} . \label{clPbot} 
\end{align}

The third set of boundary conditions results from the continuity of $B_x$, $B_y$, $B_z$ and $j_z$ at the interface between the sliding plates (\ref{Bx}), (\ref{By}), (\ref{Bz}) and (\ref{jz}). They impose continuity on $T$, $P$ and $P'$:  
\begin{align}
{T_t}_{\left| z=0 \right.} &= {T_b}_{\left| z=0 \right.} , \label{clTmid} \\
{P_t}_{\left| z=0 \right.} &= {P_b}_{\left| z=0 \right.} , \label{clPmid} \\
{P'_t}_{\left| z=0 \right.} &= {P'_b}_{\left| z=0 \right.} . \label{clPpmid} 
\end{align}

The last boundary condition is also related to the interface between the plates. It concerns the tangential components of the electric field, ${\bf E}$. From Ohm's law, ${\bf E}  = \eta {\bf j} - {\bf u} \times {\bf B}$. The induction equation states that the curl of ${\bf E}$ is equal to $-\partial {\bf B} / \partial t$, which leads to the induction equation (for conducting materials). Given that the tangential components of ${\bf E}$ are already involved in the $z$ component of the curl of ${\bf E}$, the bulk induction equation contains already partly some information on the continuity of $E_x$ and $E_y$. It is enough to impose only one other independent constraint: one possibility is to impose continuity on the horizontal divergence $k_x E_x + k_y E_y$ (independent of the $z$ curl: $k_x E_y - k_y E_x$). From (\ref{jx}) and (\ref{jy}) the horizontal divergence takes the following expression
\begin{align}
k_x E_x + k_y E_y = & i \eta _1 k_q d_q \left[ P'' -k^2 P \right] + i \left( k^2 + \eta _1 d_q^2 \right) T' \nonumber \\
& + \eta _1 q_z k^2 d_q T + k_y k^2 U P.                       \label{divE}
\end{align}
Continuity of the tangential components of the electric field can be written:
\begin{align}
&i \eta _{1t} k_{qt} d_{qt} \left[ {P''_t}_{\left| z=0 \right.} -k^2 {P_t}_{\left| z=0 \right.} \right] + i \left( k^2 + \eta _{1t} d_{qt}^2 \right) {T'_t}_{\left| z=0 \right.}  \nonumber \\
& + \eta _{1t} q_{zt} k^2 d_{qt} {T_t}_{\left| z=0 \right.} + k_y k^2 U {P_t}_{\left| z=0 \right.}   \nonumber \\
&=  i \eta _{1b} k_{qb} d_{qb} \left[ {P''_b}_{\left| z=0 \right.} -k^2 {P_b}_{\left| z=0 \right.} \right] + i \left( k^2 + \eta _{1b} d_{qb}^2 \right) {T'_b}_{\left| z=0 \right.} \nonumber \\
& + \eta _{1b} q_{zb} k^2 d_{qb} {T_t}_{\left| z=0 \right.} - k_y k^2 U {P_b}_{\left| z=0 \right.} . \label{clEtang} 
\end{align}

\section{Eigenvalue problem}

In each plate, the eigenproblem (\ref{eqP}) and (\ref{eqTnew}) can be written under the form of a matrix equation, when $P(z)$ and $T(z)$ are expanded in a series of (collocation) Chebyshev polynomials \cite{wr2000}. Their coefficients $P_i$ and $T_i$ must be solutions of 
\begin{equation} \label{eigen}
\gamma \begin{bmatrix} I & 0 \\ -\delta D & I \end{bmatrix} \begin{bmatrix} P_i \\ T_i \end{bmatrix} = \begin{bmatrix} \mathcal{Q} & \mathcal{R} \\ \mathcal{S} & \mathcal{T} \end{bmatrix} \begin{bmatrix} P_i
 \\ T_i \end{bmatrix}, 
\end{equation}
where $I$ is the identity matrix, $D$ is the first order differentiation matrix ($P'_i = D_{ij} P_j$), and the matrices $\mathcal{Q}$, $\mathcal{R}$, $\mathcal{S}$, $\mathcal{T}$ are the following
\begin{align}
\mathcal{Q} =& -i k_x U I + \left(1 + \eta_1 \frac{k_q^2}{k^2} \right) \left[ D^2 - k^2 I \right] , \label{Q} \\
\mathcal{R} =& -i \eta _1 k_q q_z I + \eta _1 \frac{k_q d_q}{k^2} D , \label{R}  \\
\mathcal{S} =& ik_x \delta U D 
 - i \eta _1 q_z k_q \left[ D^2 - k^2 I \right] ,\label{S}  \\
\mathcal{T} =& -i k_x U I + D^2 - k^2 \left( 1 + \eta _1 q_z^2 \right) I  \nonumber \\
& - i \eta _1 q_z d_q \frac{2 k^2 + \eta _1 k_q^2} {k^2 + \eta _1 k_q^2} D +  \eta _1 \frac{d_q^2 }{ k^2 + \eta _1 k_q^2} D^2 . \label{T} 
\end{align}
When the two plates are considered simultaneously, we have a global eigenvalue problem involving $P_t$, $T_t$, $P_b$ and $T_b$
\begin{equation} \label{eigenGlob}
\gamma \begin{bmatrix}\! I \!\! & 0 \!\! & 0 \!\! & 0 \! \\ \! - \delta _t D\!\! & I \!  & 0 \! \! & 0 \! \\ \! 0 \!\! & 0 \!\! & I \!\! & 0 \! \\ \! 0 \!\! &  0 \!\! & \! -\delta _b D \!\! &  I \!  \end{bmatrix} \begin{bmatrix} P_{t} \\ T_{t} \\ P_{b} \\ T_{b} \end{bmatrix} = \begin{bmatrix} \mathcal{Q}_t \!\!\! & \mathcal{R}_t \!\!\! & 0 \! & 0 \! \\ \mathcal{S}_t \!\!\! & \mathcal{T}_t \!\!\! & 0 \!\!\! & 0 \! \\ 0 \!\!\! & 0\!\! \! & \mathcal{Q}_b \!\!\! & \mathcal{R}_b \! \\ 0 \!\!\! & 0 \!\!\! & \mathcal{S}_b \!\!\! & \mathcal{T}_b \! \end{bmatrix} \begin{bmatrix} P_{t} \\ T_{t} \\ P_{b} \\ T_{b}
  \end{bmatrix}. 
\end{equation}
Top and bottom plates seem to obey independent equations, however this is going to change when the boundary conditions are taken into account. We have eight boundary conditions, meaning that eight of the components of  $P_t$, $T_t$, $P_b$ and $T_b$ are expressed in terms of the others. We have chosen each of the end Chebyshev points for $P_t$, $T_t$, $P_b$ and $T_b$. Thus, the remaining eigenvalue problem only involves the inner Chebyshev points, and the global matrices are no longer block diagonal. We symbolically write this final eigenvalue problem as follows
\begin{equation} \label{eigenFinal}
\gamma \mathcal{A} \begin{bmatrix} P_{t} \\ T_{t} \\ P_{b} \\ T_{b} \end{bmatrix} = \mathcal{B} \begin{bmatrix} P_{t} \\ T_{t} \\ P_{b} \\ T_{b} \end{bmatrix} .
\end{equation}
Under this form, the eigenvalue problem is solved using a software such as octave (free software under the GNU licence). Any eigenmode with eigenvalue of positive real part is said to be a dynamo solution. 

\section{Neutral stability}

\begin{figure}
\begin{center}
\includegraphics[width=8.5cm,keepaspectratio]{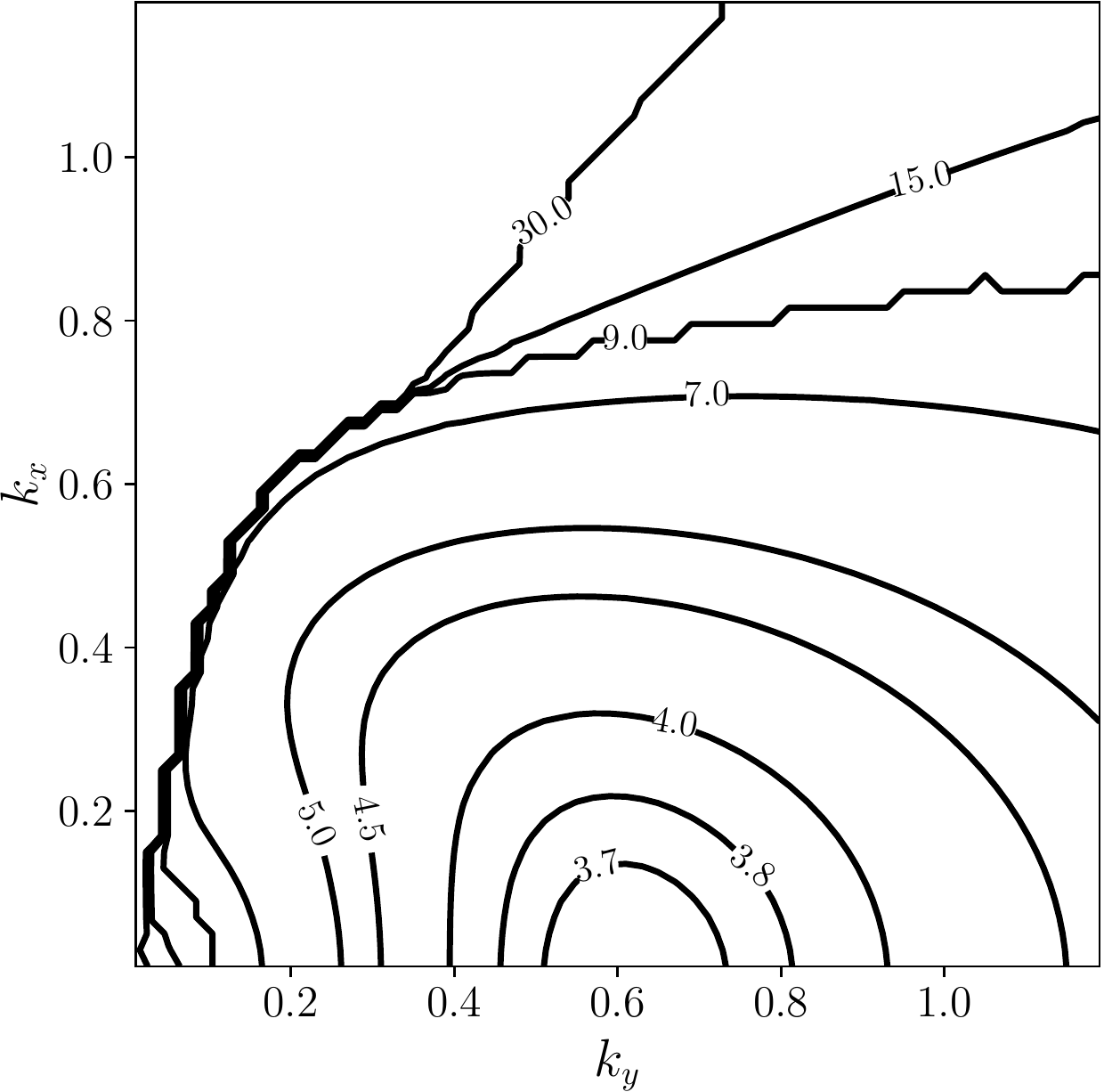}
\caption{Critical magnetic Reynolds number, for a uniform anisotropy ($\alpha = 0.5$~rad, $\beta = 0$, $\eta_1=1000$) in both plates, as a function of $k_x$ and $k_y$.}
\label{figkxky}
\end{center}
\end{figure}

As an example, we consider the case of a uniform anisotropy: $\alpha = 0.5$~rad, $\beta = 0$ and $\eta _1 = 1000$. For each value of the wavenumbers $k_x$ and $k_y$, we determine the critical magnetic Reynolds number and plot it on Fig.~\ref{figkxky}. When $k_y=0$, the critical magnetic Reynolds number is infinite. In a large domain, $0 < k_x < 0.5$ and $0.3 < k_y < 1.0$, the critical magnetic Reynolds number is less than about 5. The minimal value of the critical magnetic Reynolds number, about $3.6$, is found for $k_x=0$ and $k_y \simeq 0.62$. Fig.~\ref{figkxky} provides a justification to restrict the analysis to the case $k_x = 0$ if we are looking for minimal magnetic Reynolds numbers.

In general, the eigenvalue $\gamma$ has a non-zero imaginary part at the critical magnetic Reynolds number ({\it i.e.} $\mathrm{Re} (\gamma ) = 0$ and $\mathrm{Im} (\gamma ) \neq 0 $), however when $k_x = 0$ this is not the case since all eigenvalues are real, unless the magnetic Reynolds number is above 50 or so, well above the critical value $3.6$. For instance, for the case identified above -- $k_x=0$, $\alpha = 0.5 $~rad, $\eta _1 =1000$, $k_y=0.62$ -- we have been looking numerically for the minimal Reynolds number leading to at least one eigenvalue with non-zero imaginary part. We find that we need a magnetic Reynolds number $Rm \simeq 52.6$  to observe the first non-real eigenvalue and that its real part is about $-135$ corresponding to a strongly damped solution.

\section{Analytical solution of the neutral stability for $\lowercase{k_x}=0$ and $\beta=0$}
\label{analyt}

In the particular case of $k_x=0$ and $\beta=0$, as discussed in the previous section, the eigenvalues are real around the lowest critical magnetic Reynolds number. We thus look specifically for real eigenvalues and it is possible to derive an analytical expression for the critical magnetic Reynolds number $Rm$. 

\begin{figure}
\begin{center}
	\includegraphics[width=8.5cm,keepaspectratio]{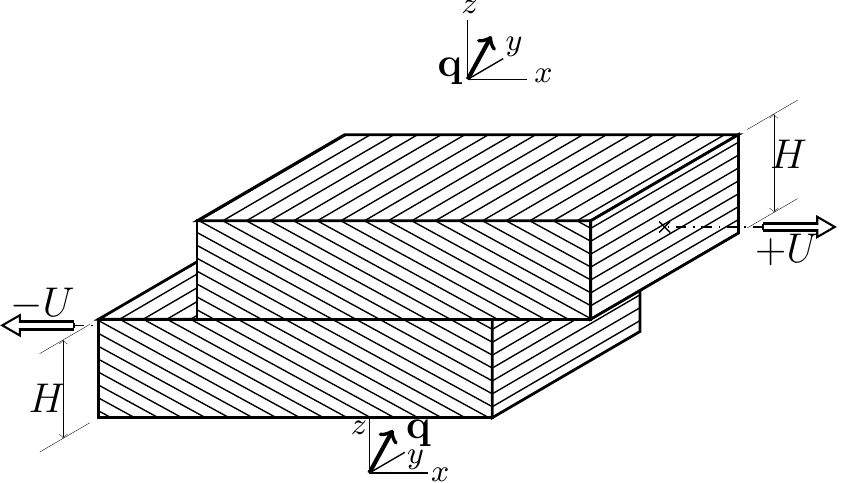}
\caption{Configuration of uniform anisotropy, with $\beta =0$, so that the direction of large electrical resistivity lies in the $(x,z)$ plane.}
\label{uniform_anisotropy}
\end{center}
\end{figure}

In this section, for simplicity, we restrict the analysis to the case of equal anisotropy in both plates (see Fig.~\ref{uniform_anisotropy}): $\alpha _t = \alpha _b = \alpha$ and $\eta _{1t} = \eta _{1b}$ (the general case is treated in appendix \ref{general_analyt}).

\begin{figure} 
\begin{center}
\includegraphics[width=8.5cm,keepaspectratio]{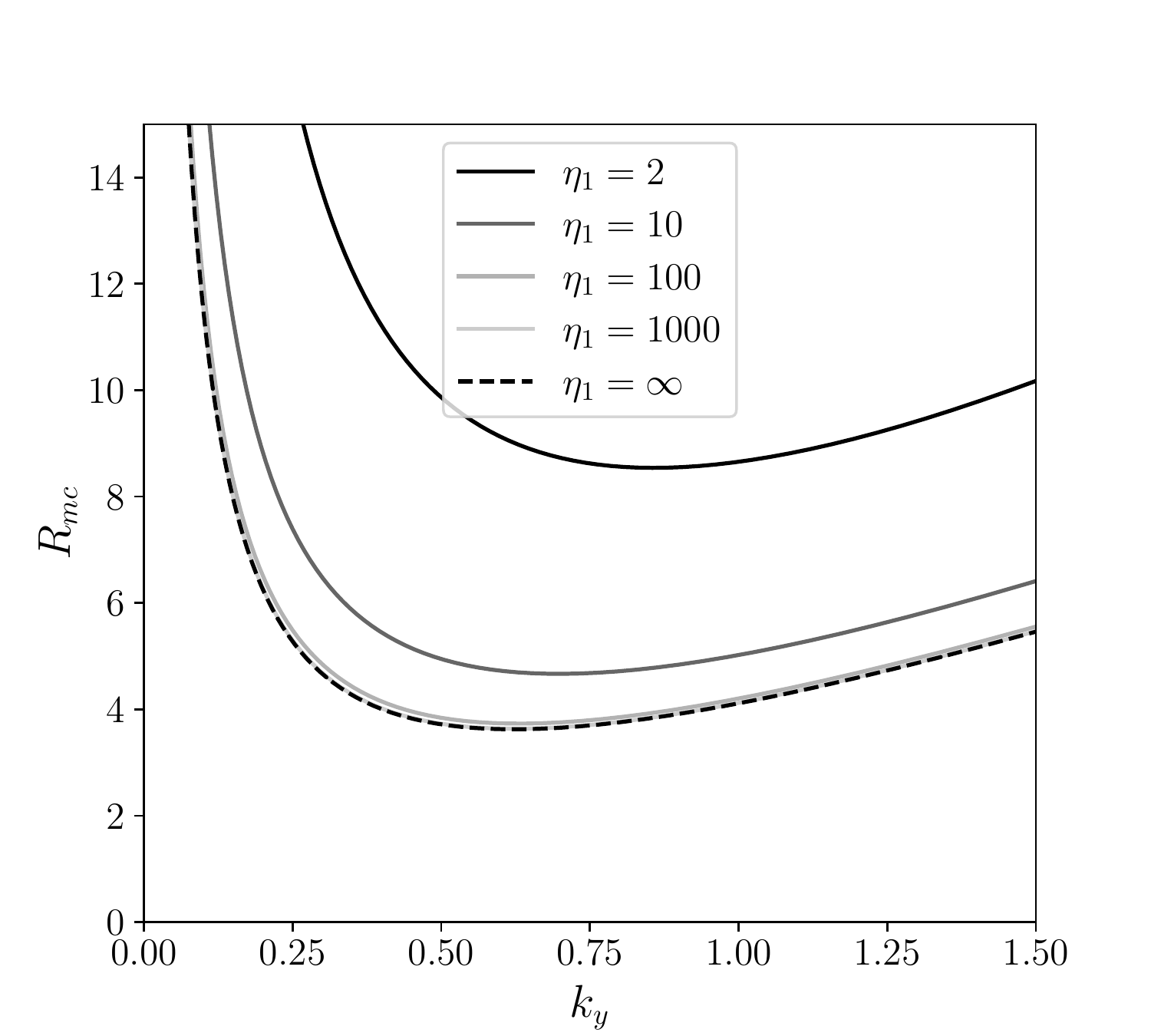}
\caption{Analytical critical magnetic Reynolds number, from equation (\ref{rmc}), for a uniform anisotropy ($\alpha = 0.5$~rad, $\beta = 0$) in both plates, as a function of $k_y$, for four values of anisotropy: $\eta _1 = 2$, $\eta _1 = 10$, $\eta _1 = 100$ and $\eta _1 = 1000$. The limiting case of infinite $\eta _1$ is also shown.}
\label{Rmcky}
\end{center}
\end{figure}

\begin{figure}
\begin{center}
\includegraphics[width=8.5cm,keepaspectratio]{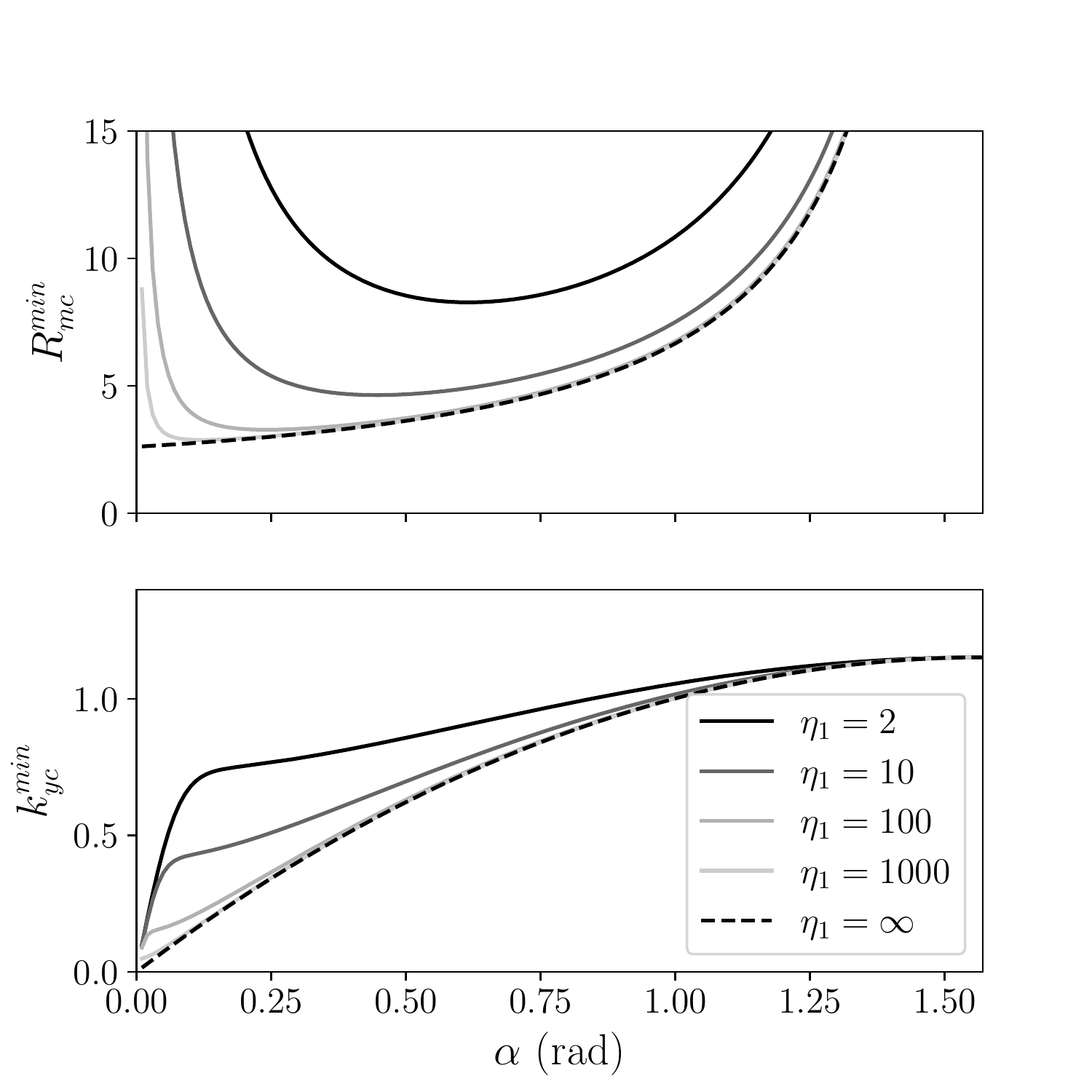}
\caption{Analytical minimal critical magnetic Reynolds number $R_{mc}^{min}$, from equation (\ref{rmc}), for a uniform anisotropy ($\beta = 0$) in both plates, as a function of $\alpha$, for four values of anisotropy: $\eta _1 = 2$, $\eta _1 = 10$, $\eta _1 = 100$ and $\eta _1 = 1000$. The limiting case of infinite $\eta _1$ is also plotted. The corresponding wavenumber $k_{yc}^{min}$ is plotted below. }
\label{minRmc}
\end{center}
\end{figure}

With those assumptions, we have $d_q = 0$ and $\delta = 0$, while $k_q=-k_y q_x$. In the critical case, $\mathrm{Re}{(\gamma)} = 0$ and since all eigenvalues are real, we also have $\mathrm{Im}{(\gamma)} = 0$, so that the critical eigenvalue equations can be written
\begin{align}
0 = & \left( 1 + \eta _1 q_x^2 \right) \left[ P'' - k_y^2 P \right]  
  + i \eta _1 k_y q_x q_z T ,  \label{eqPsimple} \\
0 = & T'' - k_y^2 T  - \eta _1 k_y^2 q_z^2 T 
 + i \eta _1 k_y q_x q_z \left[ P'' - k_y^2 P \right] . \label{eqTsimple}
\end{align}
From the first equation (\ref{eqPsimple}), we obtain an explicit expression of $T$ in terms of $P$
\begin{equation}
T = i \frac{\mathcal{F}}{k_y} \left[ P'' - k_y^2 P \right], \label{TofP}
\end{equation}
where we introduce the notation 
\begin{equation}
\mathcal{F} = \frac{\frac{1}{\eta_1} + q_x^2 }{ q_x q_z} . \label{notF}
\end{equation}
Substituting $T$ in equation (\ref{eqTsimple}) using (\ref{TofP}) leads to a fourth order ordinary differential equation, with constant coefficients, governing $P$
\begin{equation}
P'''' - [1 + \mathcal{N} ] k_y^2 P'' + \mathcal{N}  k_y^4 P = 0, \label{diffeqP}
\end{equation}
where we introduce another notation 
\begin{equation}
\mathcal{N} = \frac{ \frac{1}{\eta_1} + 1 }{\frac{1}{\eta_1} +  q_x^2} . \label{notN}
\end{equation}
The roots of the biquadratic characteristic equation associated with (\ref{diffeqP}) are easily obtained: $k_y$, $-k_y$, $\sqrt{\mathcal{N}} k_y$, $-\sqrt{\mathcal{N}} k_y$. Using the symmetry of the problem, we look for eigenvectors of the following form
\begin{align}
P_t &= a_1 e^{k_y z} + a_2 e^{-k_y z} + a_3 e^{\sqrt{\mathcal{N}}k_y z} + a_4 e^{-\sqrt{\mathcal{N}}k_y z}, \label{Pt}\\
P_b &= a_1 e^{-k_y z} + a_2 e^{k_y z} + a_3 e^{-\sqrt{\mathcal{N}}k_y z} + a_4 e^{\sqrt{\mathcal{N}}k_y z}, \label{Pb}
\end{align}
where the coefficients $a_1$, $a_2$, $a_3$ and $a_4$ have now to be determined using the boundary conditions. Note that the order of the elementary exponentials is different in (\ref{Pt}) and (\ref{Pb}) in order to ensure that the combined poloidal function $(P_t, P_b)$ is an even function of $z$. We first consider the boundary condition $T=0$ at $z=\pm 1$, which, using (\ref{TofP}), leads to
\begin{equation}
a_3 = - a_4 e^{-2\sqrt{\mathcal{N}}k_y}. \label{a3}
\end{equation}
Next, the boundary condition $k_y P + P' =0$ at $z=1$, or equivalently (by symmetry) $k_y P - P' =0$ at $z=-1$, provide
\begin{equation}
a_1 = a_4 \sqrt{\mathcal{N}} e^{-(1+\sqrt{\mathcal{N}})k_y}. \label{a1}
\end{equation}
We then need to ensure continuity of $P$ and $P'$ at the interface between the plates ($z=0$). This is automatic for $P$, as we made it a continuous even function of $z$, see (\ref{Pt}) and (\ref{Pb}), and continuity of $P'$ leads to
\begin{equation}
a_2 = a_4 \sqrt{\mathcal{N}} \left[ e^{-(1+\sqrt{\mathcal{N}})k_y} - e^{-2\sqrt{\mathcal{N}}k_y} -1 \right]. \label{a2}
\end{equation}
Continuity of $T$, given (\ref{TofP}) and the previously mentioned continuity of $P$, is equivalent to the continuity of $P''$, which is satisfied by construction for the same reason as that of $P$. There is only one last condition to consider, that of continuity of the tangent electric field (\ref{clEtang}). Now we have expressed the coefficients of the eigenvector in terms of one scalar, $a_4$. The last condition will not provide the value of this coefficient (an eigenvector can be multiplied by any non-zero scalar and is still an eigenvector!) but it will provide the condition of the existence of such an eigenvector instead. With the simplifying assumptions used in this section,  using (\ref{TofP}) and the results derived up to now (\ref{a3}), (\ref{a1}) and (\ref{a2}), equation (\ref{clEtang}) can be written as
\begin{equation}
\begin{split} \label{lastCL}
& \mathcal{F} \left( \mathcal{N} -1 \right) a_4 \left[1 + e^{-2\sqrt{\mathcal{N}}k_y} \right] + \frac{U}{k_y} a_4 \Biggl[ 2 e^{-(1+\sqrt{\mathcal{N}})k_y} \Biggr. \\ 
& \hspace*{1 cm} \left. - 1 - e^{-2\sqrt{\mathcal{N}}k_y} + \frac{1 - e^{-2\sqrt{\mathcal{N}}k_y} }{\sqrt{\mathcal{N}}}  \right] = 0 . 
\end{split}
\end{equation}
We divide equation (\ref{lastCL}) by $a_4$ and obtain the condition for the existence of a critical eigenvector. That condition is expressed as the required value of the velocity $U$ necessary to satisfy (\ref{lastCL}), which is then the critical velocity $U_c$ or critical magnetic Reynolds number\footnote{The choice of velocity scale $\left( \mu _0 \sigma _0 H \right) ^{-1}$, in section \ref{gov}, implies that the dimensionless velocity is a magnetic Reynolds number.} $R_{mc}$
\begin{equation}
R_{mc} = \frac{ k_y \mathcal{F} \left( \mathcal{N} -1 \right) \left[1 + e^{-2\sqrt{\mathcal{N}}k_y} \right]  }{1 + e^{-2\sqrt{\mathcal{N}}k_y} \left[1+\frac{1}{\sqrt{\mathcal{N}}} \right] - \frac{1}{\sqrt{\mathcal{N}}} -2 e^{-(1+\sqrt{\mathcal{N}})k_y}}. \label{rmc}
\end{equation}
Here, the critical magnetic Reynolds number $R_{mc}$ is expressed explicitly in terms of $k_y$ and the electrical anisotropy $\mathcal{F}$ and $\mathcal{N}$, who are functions of $\eta_1$, $q_x$ and $q_z$ (with the condition $q_x^2+q_z^2=1$). On figure \ref{Rmcky}, we plot $R_{mc}$ in (\ref{rmc}) as a function of $k_y$ for $\alpha = 0.5$ (corresponding to $q_x \simeq 0.47943$ and $q_z \simeq 0.87758$) and different values of $\eta _1$. 
We can see that as the ratio of resistivities 
 $\eta _1$ increases to large value, the critical curve of magnetic Reynolds
 number converges towards a limiting curve. Its expression can be derived from (\ref{rmc}), as $\mathcal{N} \longrightarrow q_z^2 / q_x^2$ and $\mathcal{F} \longrightarrow q_x / q_z$
\begin{equation}
\lim\limits_{\eta _1 \to \infty} R_{mc} = \frac{ q_z \frac{k_y}{q_x} \left[1 + e^{-2 \frac{k_y}{q_x}} \right]  }{1 + e^{-2 \frac{k_y}{q_x}} \left[1+q_x \right] - q_x -2 e^{-\left( 1+\frac{1}{q_x}\right) k_y}}. \label{rmcinf}
\end{equation}
The critical curves have always a minimal value for some $k_y$, however it is not possible to obtain its analytical expression. This is done numerically and we plot on Fig.~\ref{minRmc} the minimum critical magnetic Reynold number and corresponding wavenumber, in terms of the angle of the anisotropy $\alpha$. 
In the limit of infinite $\eta _1$, for small values of $\alpha$, the wavenumber $k_{yc}^{min}$ is proportional to $\alpha$
\begin{equation}
k_{yc}^{min} \simeq 1.505 \alpha . \label{kinf}
\end{equation}
The absolute minimum critical magnetic Reynolds number $R_{mc0}$ is obtained for infinite $\eta _1$, in the limit of vanishing $\alpha$ and its numerical value is approximately
\begin{equation}
R_{mc0} \simeq 2.609 . \label{rmc0}
\end{equation}

\begin{figure}
\begin{center}
\includegraphics[width=8.5cm,keepaspectratio]{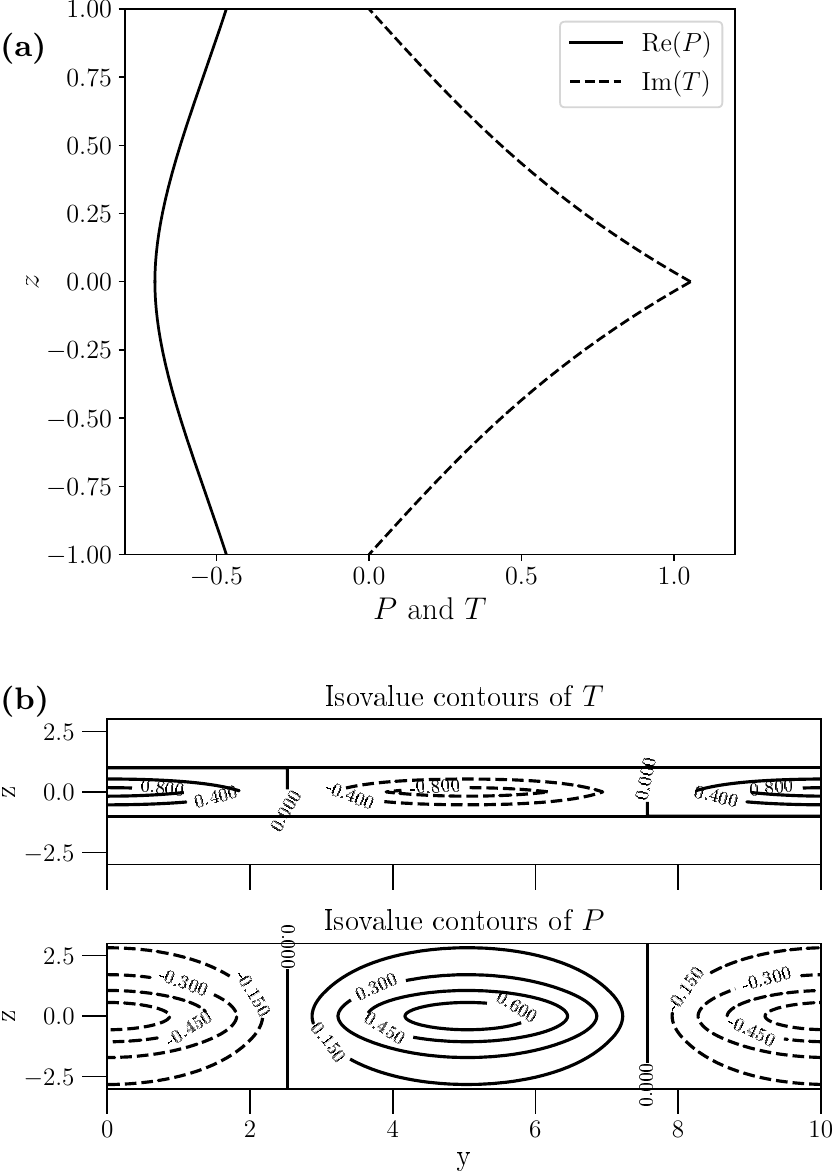}
\caption{Analytical eigenvector {\bf (a)}, both $P$ and $T$ components, for $\eta _1 = 1000$, $\alpha=0.5$, and the minimal $R_{mc} \simeq 3.6347$, attained at $k_{yc} \simeq 0.6225$. Isovalues of the eigenvector in the ($y,z)$ plane of both $T$ and $P$ components {\bf (b)} for the same parameters.}
\label{eigenvect}
\end{center}
\end{figure}

Coming back to the eigenvector itself, its analytical expression (\ref{Pt}) and (\ref{Pb}), together with (\ref{a3}), (\ref{a3}), (\ref{a1}) and (\ref{TofP}), is plotted on Fig.~\ref{eigenvect} {\bf (a)}. The corresponding contour lines in the $(y,z)$ plane are shown on Fig.~\ref{eigenvect} {\bf (b)}. The component $T$ vanishes outside the electrically conducting domain, while $P$ decays to zero at $z=\pm \infty$. Contour lines of $T$ are isolines of the $x$-component of the magnetic field and isolines of $P$ correspond to magnetic lines in the $(y,z)$ plane.

\section{Physical understanding of the analytical dynamo} 
\label{physUnder}

\begin{figure}
\begin{center}
	\includegraphics[width=8.5cm,keepaspectratio]{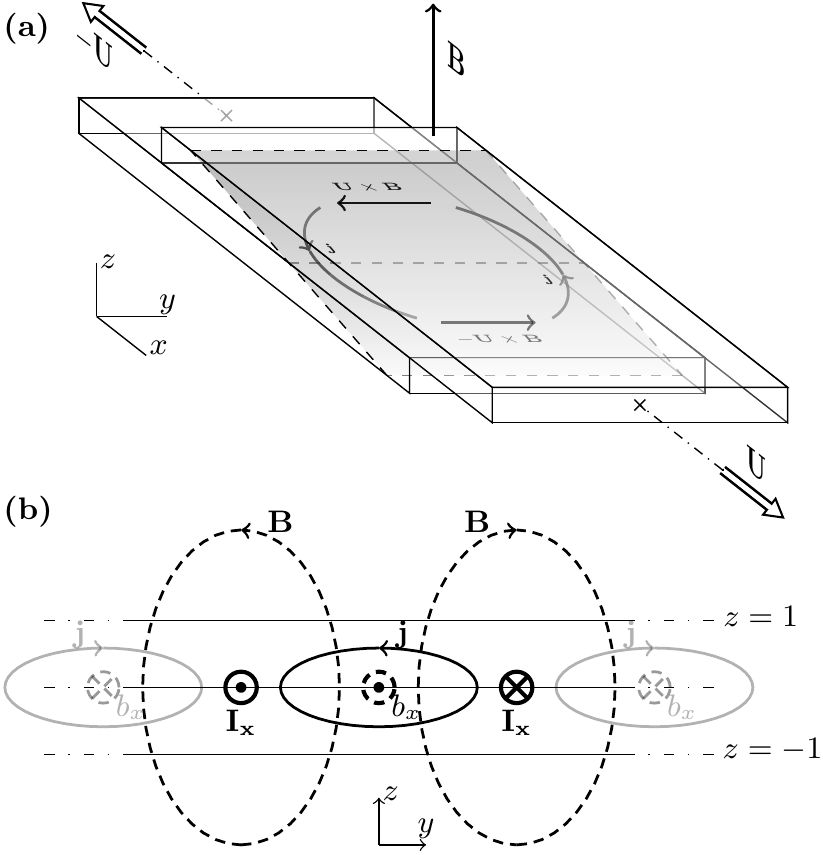}
	\caption{Same configuration as in Fig.~\ref{uniform_anisotropy}, but only one platelet of high electrical conductivity is shown in each plate {\bf (a)}. Schematic electric current circuits (thick solid lines) {\bf (b)} in the $(y,z)$ plane are labeled $I_x$ in the $x$-direction and ${\bf j}$ in the $(y,z)$-plane. The corresponding schematic magnetic field is shown (thick dashed lines), ${\bf B}$ in the $(y,z)$-plane and $b_x$ in the $x$-direction. The pattern is periodic in the $y$-direction and invariant in the $x$-direction.}
\label{Phys}
\end{center}
\end{figure}

When $\beta =0$ and for a uniform (strong) anisotropy, it is possible to use hand-waving arguments to understand how this dynamo is working. The case is shown again on Fig.~\ref{Phys} {\bf (a)}, where the sliding plates are visible and where we have made visible a single 'platelet' of large electrical conductivity (shaded area). More precisely, we have been considering those two platelets, one in each sliding plate, that happen to form a single planar surface at a given time. Let us assume that a small seed of transverse magnetic field ${\bf B}$ exists at that time. By symmetry, it is natural to consider that the transverse electric field in the $y$-direction is zero. The electric current is then driven by the induction voltage ${\bf U} \times {\bf B}$ in the upper plate and $-{\bf U} \times {\bf B}$ in the lower plate. If the transverse field ${\bf B}$ is localized, the induced electric current will generated a potential electric field and a current loop will be formed (see Figs.~\ref{Phys} {\bf (a)} and {\bf (b)}). The loop of electric current can be decomposed in a loop in the $(y,z)$ plane -- which generates a magnetic field in the $x$ direction and in longitudinal current tubes in the $x$ direction -- generating a magnetic field in the $(y,z)$ plane, in particular reinforcing the initial seed magnetic field in the $z$ direction. Those longitudinal current tubes are parallel, alternatively in the positive and negative direction. Only two of them are represented in Fig.~\ref{Phys} {\bf (b)}, which corresponds to half a period of the dynamo solution in the $y$ direction.

If the direction of the sliding plates is reversed, one can conclude with the same reasoning that the seed magnetic field is damped by the induced current loop. 

Another result of the previous section can be understood thanks to Fig.~\ref{Phys} {\bf (a)} concerning the optimal wavenumber of the critical dynamo modes. In the shaded plane on the figure, it seems natural that the electric current loop has an aspect ratio of order one, and when the angle $\alpha$ becomes small the size of the shaded area becomes larger. This is the reason why the optimum wavenumber goes to zero linearly when $\alpha$ goes to zero. In addition, when the anisotropy is not very strong (small values of $\eta _1$) the electric current 'leaks' to the neighbouring platelets instead of running over the whole area and the resulting current loops are smaller (hence the wavenumber $k_y$ is larger than in the limit $\eta _1 \rightarrow \infty$). 

In his original derivation of the impossibility to sustain an axisymmetrical (or equivalently a magnetic field invariant along one direction) by dynamo action, Cowling \cite{cowling33} uses an argument based on a neutral point in the meridional plane. This point corresponds to an extremum in the poloidal scalar, {\it i. e.} zero meridional magnetic field. For instance such at neutral point can be seen on Fig.~\ref{eigenvect} {\bf (b)} showing the contours of $P$. Its coordinates are $(x,y) \simeq (5,0)$ (and every half period $(0,0)$, $(10,0)$, ...). The argument is that there should be some electrical current flowing in the $x$ direction at a neutral point (because of the circulation of ${\bf B}$ around it) but Ohm's law makes it impossible: ${\bf u}\times {\bf B}$ is zero and there is no electric potential gradient in this direction\footnote{The argument is slightly more subtle as it considers the order of magnitude of the circulation on an infinitesimal circle around the neutral point and concludes that the electromotive force cannot match its scaling.} With the anisotropic  electrical conductivity considered in the present work, it is the electric potential gradient in the $z$ direction that causes an electric current to flow in the $x$ direction. 

\section{Growth rate of the dynamo}
\label{gamma}

\begin{figure}
\begin{center}
\includegraphics[width=8.5cm,keepaspectratio]{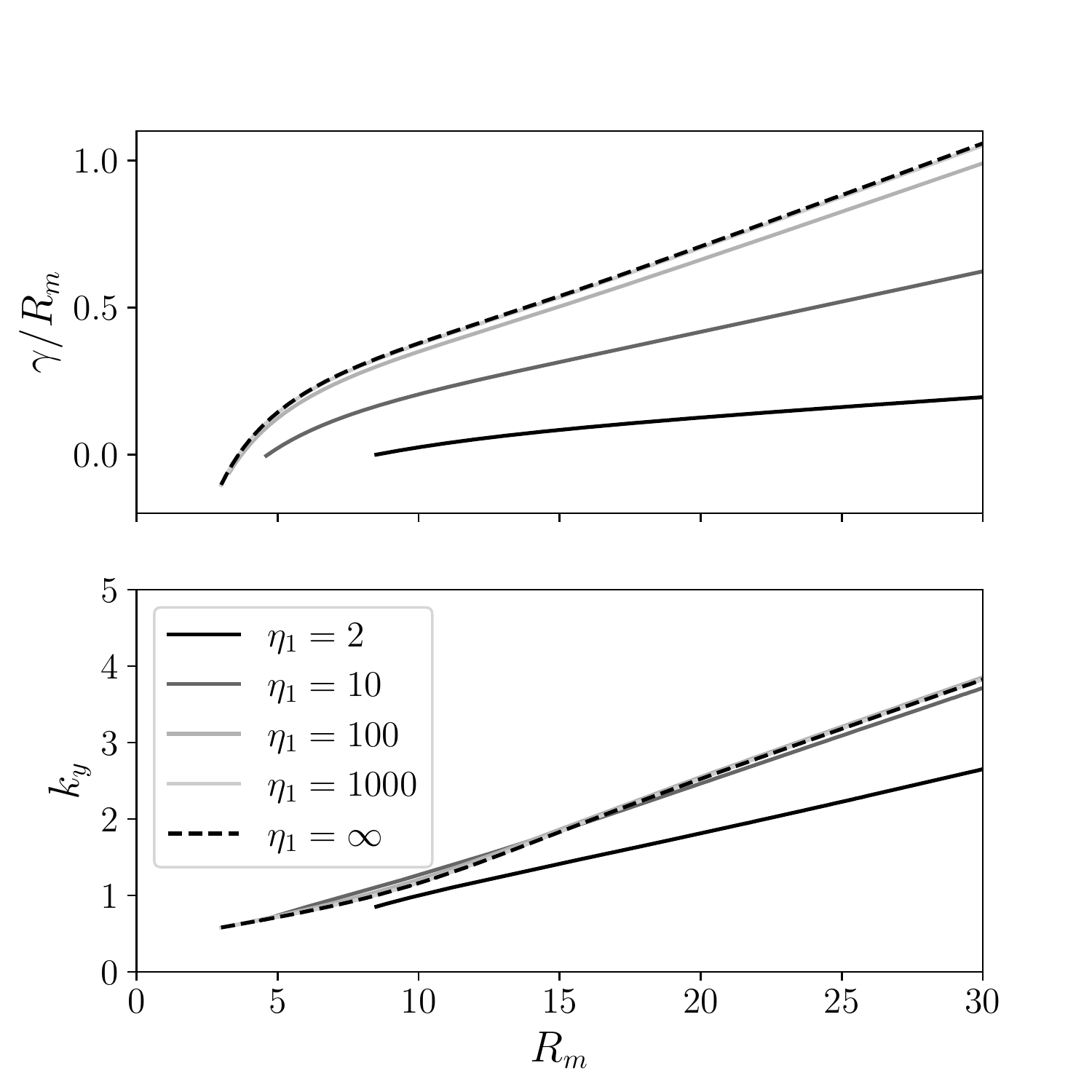}
\caption{Optimal growth rate and corresponding wavanumber $k_y$, for a uniform anisotropy ($\alpha = 0.5$~rad, $\beta = 0$) in both plates, as a function of $R_m$, for five values of anisotropy: $\eta _1 = 2$, $\eta _1 = 10$, $\eta _1 = 100$, $\eta _1 = 1000$ and $\eta _1 = \infty$. Note that the growth rate is divided by $R_m$, which corresponds to the dimensional growth rate divided by $U/H$.}
\label{fig9}
\end{center}
\end{figure}

It is also possible to derive an analytical expression for the (real) growth rate of this dynamo for supercritical values of the magnetic Reynolds number. The derivation is done explicitly, see equation (\ref{Rmga}) in appendix \ref{rate}, in the case $k_x = 0$, $\eta _{1t} = \eta _{1b}$, $q_{yt}=q_{yb}=0$ and $q_{xt}=q_{xb}$ (hence $q_{zt}=q_{zb}$). We have investigated the optimal growth rate and corresponding optimal wavenumber for the particular angle $\alpha =0.5$~rad of anisotropic orientation ($\beta = 0$). Using equation (\ref{Rmga}), we find numerically for each value of $R_m$ the maximal growth rate $\gamma$ and the corresponding wavenumber $k_y$ and plot them in Fig.~\ref{fig9}. For simplicity, we have restricted the plot to the condition $\gamma > -k_y^2$, so that the eigenvectors are simple real exponentials and equation (\ref{Rmga}) can be inverted to express $\gamma$ as a function of $R_m$. Above $R_m \simeq 15$ or so, the optimal growth rate (made dimensionless using $U/H$, exactly equal to $\gamma / R_m$) and corresponding wavenumber increase linearly with $R_m$. The slope of the optimal growth rate depends on the degree of anisotropy $\eta _1$. 

For asymptotic large values of $Rm$, we obtain asymptotic expressions for the fastest growth rate and corresponding wavenumber (see appendix \ref{rate}). we find that, as $Rm$ is increased, the dimensionless growth rate increases proportionally to $Rm^2$. Back to dimensional values, this corresponds to a growth rate proportional to $Rm U/H$, hence increasing linearly with the electrical conductivity. In the terminology introduced by Vainshtein and Zeldovich \cite{VZ72}, a 'fast' dynamo is such that its growth rate remains finite as $Rm$ (or electrical conductivity) goes to infinity. Thus we suggest that our dynamo could be called a 'very fast' dynamo, with its increasing, unbounded, growth rate with $Rm$ (or the electrical conductivity). In contrast, the uniform shear flow with anisotropic electrical conductivity (\cite{RR84}) leads to a 'fast' dynamo, not 'very fast'. The derivation of those scalings is shown in appendix \ref{growth} using a scaling analysis of the governing equations.

The growth rate of a dynamo mechanism, whether it is slow, fast or very fast, is particularly important in the astrophysical and interstellar context, where the magnetic Reynolds numbers are very large. If we consider a galaxy, it is tempting to imagine that the electric conductivity is anisotropic due to the presence of filaments (see typical spiral galaxies). Differential Keplerian rotation and spiralling anisotropy of the electrical conductivity are enough, in principle, to trigger "fast" dynamo action. The solar tachocline might also be a place of application of our dynamo model: shear localization is present, however it is not so obvious why electrical conductivity should be anisotropic.

\section{Conclusions}
\label{conclusion}

In this paper, we have presented an elementary dynamo solution for an anisotropic electrical conductivity. The velocity field has a very simple structure, consisting of a solid plate sliding on another solid plate. The plates must be in electrical contact. In the dynamo terminology, this is an homogeneous dynamo, when the anisotropy is identical in both plates. With an isotropic material, we know that the velocity field must be rather complex, and cannot be a planar flow. Introducing some small complexity in the materials properties -- the electrical conductivity anisotropy -- allows us to obtain a dynamo effect with a simple planar flow. Importantly, we have shown that when the anisotropy is located witin the shear plane direction, the lowest critical magnetic Reynolds number for dynamo action is obtained for modes invariant along the velocity direction. In these cases, we obtain an analytical expression for the critical magnetic Reynolds number. Moreover, the analysis is elementary and involves only exponential solutions. It is then well suited for teaching purpose. Let us note also, that this configuration is quite an efficient dynamo, as the critical magnetic Reynolds number is low, with a minimum around $2.6$ based on the thickness of one plate and half the velocity difference, or $10.4$ based on the thickness of both plates and total velocity difference. In the supercritical regime, we also obtain an analytical expression for the growth rate, showing that the dynamo behaves as a 'very fast' dynamo: its growth rate increases linearly with electrical conductivity when all other dimensional parameters are kept constant. Last but not least, the dynamo mechanism has been analyzed and it is straigthforward to understand the electric current path which is responsible for sustaining the magnetic field.

Instead of the effect of the anisotropy of electrical conductivity, one might also consider the effect of the anisotropy of magnetic permeability. Both cases are not mathematically equivalent, but one may anticipate that this would lead to the possibility of obtaining simple dynamos with simple velocity fields. A case of heterogeneous magnetic permeability has been studied in \cite{gallet2013}, who refer to a previous study on heterogeneous electrical conductivity \cite{BW92}. Heterogeneity and anisotropy are different properties, however they are somewhat connected: heterogeneity at small scale is sometimes treated macroscopically under the form of large-scale anisotropy. For instance, the distribution of fractures in rocks at small scales has been represented as a medium with anisotropic elastic properties in \cite{ss95}, because the distribution of heterogeneities is itself anisotropic. Many other anisotropic properties emerge from the theory of 'homogenization' or 'effective medium' in various branches of physics. 

Among the previous fluid dynamo experiments, the VKS experiment \cite{cadarache} bears some similarities to our present dynamo model. The spiralling blades of the rotating disks may correspond to some degree of anisotropy of the electrical conductivity and there is of course some differential shear velocity between the bulk of the fluid and the region near the disks. We should not forget however that the magnetic permeability of the disks plays a role in that experiment (see \cite{gallet2013}), so that electrical conductivity cannot be the sole parameter to consider. For instance, dynamo action has been obtained in VKS with different configurations of iron blades (see Table I in \cite{miralles13}). In the usual 'non-scooping' direction, the critical magnetic Reynolds number is reported to be equal to 44, while it becomes 60 for straight blades (purely radial anisotropy) and reaches 68 when the curved blades rotate in the 'scooping' direction. 

The configuration we have studied can be tested experimentally. This might be possible with flat composite plates, made of thin layers of copper and insulating material alternatively. It is perhaps easier to consider a variation of the configuration, which consists in bending the plates in the direction of the flow and change them into co-axial cylinders, in differential rotation, and still in electrical contact. This axisymmetric dynamo is studied in a companion paper \cite{PA2019}. 

Our dynamo is somewhat an intermediate device between experimental homogeneous dynamos and electrical machines, such as the dynamo built by Siemens \cite{siemens} and other inventors. We can then envisage a class of configurations between our solid dynamo and a purely fluid Couette or Taylor-Couette dynamo. In the present configuration, the electrical contact between the two plates can be done using a film of liquid metal. Suppose that the thickness of the fluid is increased, the configuration will then evolve from the solid dynamo presented in this paper to a more complex -- potentially non-linear -- situation where the fluid electromotive force will come into play. The attempts to produce experimental fluid dynamos have been sometimes successful, with the Riga \cite{riga}, Karlsruhe \cite{karlsruhe} and Cadarache \cite{cadarache} dynamos, but were more often unsuccessful. Using the strategy described above, one should necessarily have dynamo action for a thin fluid layer and still have it up to a certain thickness depending presumably on the capabilities of the setup in terms of velocity or power available. 

\section*{Acknowledgments}

This work was supported by the Programme National de Plan\'etologie (PNP) of CNRS/INSU, co-funded by CNES.


\bibliography{bib}


\appendix

\section{Derivation of the scalar equations for $P$ and $T$}
\label{derivationPT}

From the poloidal-toroidal decomposition (\ref{PoloidalToroidal}) and the form of the eigenvectors (\ref{Peig}) and (\ref{Teig}), we obtain the following $(x,y,z)$ components for the magnetic fields
\begin{align}
B_x &= i k_x P' + i k_y T, \label{Bx} \\
B_y &= i k_y P' - i k_x T,  \label{By}\\
B_z &= k^2 P. \label{Bz}
\end{align}
Its curl is the dimensionless electric current density ${\bf j} = {\bnabla} \times {\bf B}$
\begin{align}
j_x &= i k_x T' - i k_y \left[ P'' - k^2 P \right] , \label{jx} \\
j_y &= i k_y T' + i k_x \left[ P'' - k^2 P \right] , \label{jy} \\
j_z &= k^2 T. \label{jz}
\end{align}
When multiplied by the (dimensionless) anisotropic diffusivity tensor (\ref{diffusivity}), we obtain
\begin{align}
\left[ \eta {\bf j} \right] _x = & i \left( \eta _1 q_x d_q + k_x \right) T' + i \left( \eta _1 q_x k_q - k_y \right) \left[ P'' - k^2 P \right]  \nonumber \\
 & + \eta _1 q_x q_z k^2 T ,    \\
\left[ \eta {\bf j} \right] _y = & i \left( \eta _1 q_y d_q + k_y \right) T' + i \left( \eta _1 q_y k_q + k_x \right)  \left[ P'' - k^2 P \right]  \nonumber \\
& + \eta _1 q_y q_z k^2 T , \\
\left[ \eta {\bf j} \right] _z = & i \eta _1 q_z \left( d_q T' + k_q \left[ P'' - k^2 P \right] \right) \nonumber \\
& + \left( \eta _1 q_z^2 + 1 \right) k^2 T .
\end{align}
Its curl is the last term in the induction equation (\ref{induction})
\begin{align}
\left[ \bnabla \times \eta {\bf j} \right] _x & =  -i \left( \eta _1 q_y d_q + k_y \right) T'' - \eta _1 \left( k^2 q_y + k_y d_q \right) q_z T'  \nonumber \\
  - & \eta _1 k_y k_q q_z \left[ P'' - k^2 P \right] + i k_y k^2 \left( \eta _1 q_z^2 + 1 \right) T    \nonumber   \\
 - & i \left( \eta _1 q_y k_q + k_x \right)  \left[ P''' - k^2 P' \right] ,  \\
\left[ \bnabla \times \eta {\bf j} \right] _y & = i \left( \eta _1 q_x d_q + k_x \right) T'' + \eta _1 \left( k^2 q_x + k_x d_q \right) q_z T'         \nonumber \\
  + & \eta _1 k_x k_q q_z \left[ P'' - k^2 P \right] - i k_x k^2 \left( \eta _1 q_z^2 + 1 \right) T    \nonumber   \\
 + & i \left( \eta _1 q_x k_q - k_y \right)  \left[ P''' - k^2 P' \right] ,  \\
\left[ \bnabla \times \eta {\bf j} \right] _z & =  - \eta _1 k_q d_q T' - \left( k^2 + \eta _1 k_q^2 \right) \left[ P'' - k^2 P \right]         \nonumber \\
 + & i \eta _1 k_q q_z T . \label{cdiffz}
\end{align}
We need also the $z$ component   of the curl of the previous vector
\begin{align}
\left[ \bnabla \times  \bnabla \times \eta {\bf j} \right] _z & = \left(1 + \eta _1 q_z^2 \right) k^4 T + i  \eta _1 q_z k_q k^2 \left[ P'' - k^2 P \right]    \nonumber \\
- & \eta _1 k_q d_q \left[ P''' - k^2 P' \right] + 2 i \eta _1 d_q q_z k^2 T' \nonumber \\
- & \left( k^2 + \eta _1 d_q^2 \right) T''. \label{ccdiffz}
\end{align}
Let us now consider the electromotive force in the induction equation. The velocity field, in each plate, consists in a uniform velocity $U$ in the $x$ direction
\begin{align}
u_x &= U , \\
u_y &= 0, \\
u_z &= 0, 
\end{align}
From the magnetic field (\ref{Bx}), (\ref{By}) and (\ref{Bz}), we compute the electromotive force ${\bf u} \times {\bf B}$
\begin{align}
\left[ {\bf u} \times {\bf B} \right] _x &= 0, \\
\left[ {\bf u} \times {\bf B} \right] _y &= - k^2 U P, \\
\left[ {\bf u} \times {\bf B} \right] _z &= i k_y U P' - i k_x U T .
\end{align} 
Its curl appears in the induction equation
\begin{align}
	\left[\bnabla \times \left( {\bf u} \times {\bf B} \right) \right] _x &= k_x^2 U P' + k_x k_y U T , \label{cUBx} \\
	\left[\bnabla \times \left( {\bf u} \times {\bf B} \right) \right] _y &= k_x k_y U P' - k_x^2 U T , \label{cUBy} \\
\left[\bnabla \times \left( {\bf u} \times {\bf B} \right) \right] _z &= -i k_x k^2 U P . \label{cUBz}
\end{align}
Finally, we need the $z$ component of the curl of the previous vector field
\begin{equation}
\left[\bnabla \times \bnabla \times \left( {\bf u} \times {\bf B} \right) \right] _z = - i k_x k^2 U  T . \label{ccUBz}
\end{equation}
We now have all parts of equations (\ref{eqP}) and (\ref{eqT}). Those equations correspond to the $z$ component of the induction equation and its curl, both divided by $k^2$. On the left-hand side of (\ref{eqP}), the time-derivative of $B_z$ is $\gamma$ times (\ref{Bz}), divided by $k^2$. On the right-hand side, we have the term in (\ref{cdiffz}) for magnetic diffusion and (\ref{cUBz}) for the electromotive part (all terms divided by $k^2$). Concerning equation (\ref{eqT}), the left-hand side is similarly obtained from (\ref{jz}) and the right-hand side from (\ref{ccdiffz}) and (\ref{ccUBz}). Again these terms are divided by $k^2$ in (\ref{eqT}). 

\section{General analytical solution for $\lowercase{k_x}=0$ and $\beta = 0$}
\label{general_analyt}

We present here a more general case than in section \ref{analyt}, where the angle $\alpha$ and degree of anisotropy $\eta _1$ differ in each plate. The angle of anisotropic direction is denoted $\alpha _t$ and $\alpha _b$ in the top and bottom plate respectively, while the degree of anisotropy is $\eta _{1t}$ and $\eta _{1b}$ respectively. Both $\alpha$ and $\eta _1$ are uniform within each plate, and the angle $\beta$ and wavenumber $k_x$ are still taken to be zero. In that case, we still observe that the eigenvalues are real in the numerical results, unless we consider large magnetic Reynolds numbers (above 50) and have some eigenvalues with a non-zero imaginary component. We look for the critical magnetic Reynolds (zero real part of the eigenvalue) assuming the imaginary part is zero. In each plate, equation (\ref{diffeqP}) is valid, with $\mathcal{N}$ defined in equation (\ref{notN}). We now have critical an eigenvector $P_t$ (resp. $P_b$) and $\mathcal{N}_t$ (resp. $\mathcal{N}_b$) in the top (resp. bottom) plate. Hence equations (\ref{Pt}) and (\ref{Pb}) are replaced by 
\begin{align}
P_t &= a_1 e^{k_y z} + a_2 e^{-k_y z} + a_3 e^{\sqrt{\mathcal{N}_t}k_y z} + a_4 e^{-\sqrt{\mathcal{N}_t}k_y z}, \label{Ptg}\\
P_b &= b_1 e^{-k_y z} + b_2 e^{k_y z} + b_3 e^{-\sqrt{\mathcal{N}_b}k_y z} + b_4 e^{\sqrt{\mathcal{N}_b}k_y z}, \label{Pbg}
\end{align}
with 
\begin{equation}
	\mathcal{N}_t = \frac{ \frac{1}{\eta_{1t}} + 1 }{\frac{1}{\eta_{1t}} +  q_{xt}^2} \hspace*{2 cm} \mathcal{N}_b = \frac{ \frac{1}{\eta_{1b}} + 1 }{\frac{1}{\eta_{1b}} +  q_{xb}^2}. \label{notNtb}
\end{equation}
In each plate, the toroidal component $T$ can still be obtained from $P$ through equation (\ref{TofP}). We now have
\begin{equation}
	T_t = i \frac{\mathcal{F}_t}{k_y} \left[ P''_t - k_y^2 P_t \right], \hspace*{1 cm} T_b = i \frac{\mathcal{F}_b}{k_y} \left[ P''_b - k_y^2 P_b \right],   \label{TofPtb}
\end{equation}
where we now have 
\begin{equation}
	\mathcal{F}_t = \frac{\frac{1}{\eta_{1t}} + q_{xt}^2 }{ q_{xt} q_{zt}} , \hspace*{1 cm} \mathcal{F}_b = \frac{\frac{1}{\eta_{1b}} + q_{xb}^2 }{ q_{xb} q_{zb}} . \label{notFtb}
\end{equation}
Again, applying the boundary conditions will lead us to obtain the critical magnetic Reynolds number. The boundary condition $T=0$ at $z=\pm 1$, using (\ref{TofPtb}), leads to
\begin{equation}
	a_4 = - a_3 e^{2 k_y \sqrt{\mathcal{N}_t}}, \hspace*{1 cm}b_4 = - b_3 e^{2 k_y \sqrt{\mathcal{N}_b}}. \label{ab34}
\end{equation}
The conditions $k_y P_t+ P'_t =0$ at $z=1$ and $k_y P_b - P'_b =0$ at $z=-1$, provide
\begin{equation}
	a_1 = - a_3 \sqrt{\mathcal{N}_t} e^{k_y (\sqrt{\mathcal{N}_t-1})}, \hspace*{0.2 cm} b_1 = - b_3 \sqrt{\mathcal{N}_b} e^{k_y (\sqrt{\mathcal{N}_b-1})}.\label{ab13}
\end{equation}\\
\noindent
Continuity of $P$ at $z=0$ ($P_t (0) = P_b (0)$), continuity of $P'$ ($P'_t (0) = P'_b (0)$) and continuity of $T$ ($T_t (0) = T_b (0)$, using (\ref{TofPtb})) can be written
\begin{align}
a_1 + a_2 + a_3 + a_4 &= b_1 + b_2 + b_3 + b_4, \label{contPtb} \\
a_1 - a_2 + \sqrt{\mathcal{N}_t} (a_3 - a_4) &= - b_1 + b_2 - \sqrt{\mathcal{N}_b} (b_3 - b_4), \label{contPprimetb} \\
\mathcal{F}_t \left( \mathcal{N}_t -1  \right) (a_3 - a_4 ) &=  \mathcal{F}_b \left( \mathcal{N}_b -1  \right) (b_3 - b_4 ) . \label{contTtb}
\end{align}
Using (\ref{ab34}), equation (\ref{contTtb}) provides
\begin{align}
a_3 \mathcal{F}_t \left( 1 - e^{2 k_y \sqrt{\mathcal{N}_t}} \right)  \left( \mathcal{N}_t -1  \right) = & \nonumber \\
	b_3 \mathcal{F}_b \left( 1 - e^{2 k_y \sqrt{\mathcal{N}_b}} \right) & \left( \mathcal{N}_b -1  \right) , \label{a3b3}
\end{align}
allowing us to express $b_3$ in terms of $a_3$.
The sum of (\ref{contPtb}) and (\ref{contPprimetb}) leads to
\begin{align}
	b_2 = a_1 & + a_3 \frac{\sqrt{\mathcal{N}_t} + 1 }{2} + a_4 \frac{- \sqrt{\mathcal{N}_t} + 1 }{2}  \nonumber \\
	& - b_3  \frac{- \sqrt{\mathcal{N}_b} + 1 }{2}  -b_4 \frac{ \sqrt{\mathcal{N}_b} + 1 }{2}, \label{b2tb}
\end{align}
while the difference between (\ref{contPtb}) and (\ref{contPprimetb}) provides
\begin{align}
	a_2 = b_1 & - a_3 \frac{1 - \sqrt{\mathcal{N}_t}}{2} - a_4 \frac{1 + \sqrt{\mathcal{N}_t} }{2} \nonumber \\
	&	+ b_3 \frac{1 + \sqrt{\mathcal{N}_b} }{2} + b_4 \frac{1 - \sqrt{\mathcal{N}_b} }{2}, \label{a2tb}
\end{align}
We now have expressed all coefficients $a_i$ and $b_i$ in terms of a single of them $a_3$. We have one last boundary condition to consider, related to the continuity of the tangential electric field (\ref{clEtang}), which can be written as $i T'_t (0) + k_y U P_t (0) = i T'_b {0} - k_y U P_b (0)$. Using (\ref{Ptg}) and (\ref{Pbg}), this condition leads to
\begin{align}
	& - \mathcal{F}_t \sqrt{\mathcal{N}_t} \left( \mathcal{N}_t -1 \right) (a_3 -a_4 ) + 2 \frac{U}{k_y} ( a_1 + a_2 + a_3 + a_4 )  \nonumber  \\
	&\hspace*{1 cm} = \mathcal{F}_b \sqrt{\mathcal{N}_b} \left( \mathcal{N}_b -1 \right) (b_3 -b_4 ). \label{Etangtb}
\end{align}
Substituting all variables $a_i$ and $b_i$ in terms of $a_3$, and making $a_3 = 1$ arbitrarily because an eigenvector is defined up to a multiplicative factor, we obtain an explicit expression for the velocity $U$ which is the value of the critical magnetic Reynolds number $R_{mc}$ 
\onecolumngrid
\begin{equation}
	R_{mc} = \frac{\mathcal{F}_t \mathcal{F}_b \left( \mathcal{N}_t -1 \right) \left( \mathcal{N}_b -1 \right) \left[ \left( 1 + e^{2 k_y \sqrt{\mathcal{N}_t}} \right)  \left( 1 - e^{2 k_y \sqrt{\mathcal{N}_b}} \right) \sqrt{\mathcal{N}_t} + \left( 1 + e^{2 k_y \sqrt{\mathcal{N}_b}} \right)  \left( 1 - e^{2 k_y \sqrt{\mathcal{N}_t}} \right) \sqrt{\mathcal{N}_b} \right] }{ f(k_y, \mathcal{N}_t ) \mathcal{F}_b \left( \mathcal{N}_b -1 \right) \left( 1 - e^{2 k_y \sqrt{\mathcal{N}_b}} \right) + f(k_y, \mathcal{N}_b ) \mathcal{F}_t \left( \mathcal{N}_t -1 \right) \left( 1 - e^{2 k_y \sqrt{\mathcal{N}_t}} \right)} \frac{k_y}{2}, \label{Rmcgen}
\end{equation}
where the function $f$ is defined as
\begin{equation}
	f(k_y , \mathcal{N} ) = - \sqrt{\mathcal{N} } e^{k_y \left( \sqrt{\mathcal{N}} -1 \right)} + \frac{\sqrt{\mathcal{N}} + 1 }{2} + \frac{\sqrt{\mathcal{N}} - 1 }{2} e^{2 k_y \sqrt{\mathcal{N}}}. \label{funcf}
\end{equation}
\twocolumngrid
It can be checked that the above expression (\ref{Rmcgen}) becomes exactly (\ref{rmc}) when anisotropy is identical in both plates: $\mathcal{N}_t = \mathcal{N}_b$, $\mathcal{F}_t = \mathcal{F}_b$. 

\section{Growth rate of dynamo modes}
\label{rate}

Let us now consider the growth rate of the dynamo modes when the magnetic Reynolds number exceeds the value corresponding to the neutral stability. We may also consider lower values of the magnetic Reynolds number, but this case is less interesting, and potentially more difficult to solve as will be seen later. 
The simplest case of uniform anisotropy is treated here, with $\eta _{1t} = \eta _{1b} = \eta _1$, $\alpha _{t} = \alpha _{b} = \alpha$, $\beta _{t} = \beta _b = 0$ and $k_x=0$. 
We look for purely real eigenvalues, $\gamma$, since the numerical solutions show that the fastest growing modes possess real eigenvalues indeed, under the assumptions mentioned above. 
Equations (\ref{eqPsimple}) and (\ref{eqTsimple}) become 
\begin{align}
\gamma P = & \left( 1 + \eta _1 q_x^2 \right) \left[ P'' - k_y^2 P \right]
  + i \eta _1 k_y q_x q_z T ,  \label{eqPsimplega} \\
\gamma T = & T'' - k_y^2 T  - \eta _1 k_y^2 q_z^2 T
 + i \eta _1 k_y q_x q_z \left[ P'' - k_y^2 P \right] . \label{eqTsimplega}
\end{align}
From (\ref{eqPsimplega}), we obtain $T$ in terms of $P$
\begin{equation}
	T = i \frac{\mathcal{F}}{k_y} \left[ P'' - k_y^2 P \right] - i \frac{\gamma}{k_y \eta _1 q_x q_z} P, \label{TofPga}
\end{equation}
where $\mathcal{F}$ is defined in equation (\ref{notF}). Substituting $T$ in (\ref{eqTsimplega}), using (\ref{TofPga}) leads to 
\begin{align}
	& P'''' \left( 1 + \eta _1 q_x^2 \right) - k_y^2 P'' \left[\frac{\gamma}{k_y^2} ( 2 + \eta _1 q_x^2 ) +2 + \eta _1 (1+ q_x^2 )  \right] \nonumber \\
	&+ k_y^4 P \left[ \left(\frac{\gamma }{k_y^2} + 1 \right) ^2 + \eta _1 \left( \frac{\gamma }{k_y^2} + 1 \right)   \right] = 0 .  \label{diffPga}
\end{align}
We look for elementary solutions under the form $e^{rz}$ to this linear differential equation with constant coefficients and obtain four solutions $r=r_1$, $r=-r_1$, $r=r_3$ and $r=-r_3$, with
\begin{align}
	r_1^2 &= \gamma + k^2, \label{r1sq} \\
	r_3^2 &= \frac{\gamma + (1 + \eta _1 ) k^2}{1+ \eta _1 q_x^2}. \label{r3sq} 
\end{align}
When we consider real positive growth rates $\gamma$, both $r_1^2$ and $r_3^2$ are real positive, so that we make the assumption that $r_1$ and $r_3$ are real and positive. If the gowth rate is slightly negative, this may still be the case. Obviously the case of strongly negative values of $\gamma$ will make $r_1^2$ and $r_3^2$ negative and the solution swill no longer be real exponentials. We shall not consider the latter case. The function $P$ is expanded as follows 
\begin{equation}
P = a_1 e^{r_1 z} + a_2 e^{-r_1 z} + a_3 e^{r_3 z} + a_4 e^{-r_3 z}, \label{Pga}
\end{equation}
in the upper plate, while it is supposed to be symmetrical in the lower plate (so that $P$ is an even function of $z$). From (\ref{TofPga}), we have
\begin{equation}
	T = k \left[ - a_1 \tilde{\gamma} e^{r_1 z} + a_2 \tilde{\gamma} e^{-r_1 z} - a_3 \frac{q_z}{q_x} e^{r_3 z} - a_4 \frac{q_z}{q_x} e^{-r_3 z} \right], \label{Tga}
\end{equation}
where 
\begin{equation}
	\tilde{\gamma} = \frac{\gamma q_x}{k_y^2 q_z}. \label{tildega} 
\end{equation}
We consider the same boundary conditions as in section (\ref{analyt}), {\it i.e.} that $P'$ is zero at $z=0$, that $T'=-k_y U P$ at $z=0$, that $T=0$ at $z=1$ and $P'+k_y P = 0$ at $z=1$
\begin{align}
	r_1 a_1 -r_1 a_2 + r_3 a_3 - r_3 a_4 & = 0 , \label{cl1} \\
	 \tilde{\gamma} r_1 (a_2-a_1)   - \frac{q_z}{q_x} r_3 (a_3 - a_4) \hspace{1 cm} & \nonumber \\
	 + U (a_1 + a_2 + a_3 + a_4 ) &= 0 , \label{cl2} \\
	 \tilde{\gamma} e^{r_1} a_1 + \tilde{\gamma} e^{-r_1} a_2 + \frac{q_z}{q_x} e^{r_3} a_3 + \frac{q_z}{q_x} e^{-r_3} a_4 & = 0 , \label{cl3} \\
	 (k_y + r_1) e^{r_1} a_1 + (k_y -r_1) e^{-r_1} a_2  \hspace{1 cm} & \nonumber \\ 
	 + (k_y + r_3) e^{r_3} a_3 + (k_y -r_3) e^{-r_3} a_4 & = 0 . \label{cl4} 
\end{align}
The system above has a non-trivial solution when the determinant of the underlying matrix is zero, which is the condition for the existence of an eigenvalue. So the equation relating $U$ (which is also $Rm$ as discussed above and will be denoted as such in the following) and the governing parameters (including here the growth rate $\gamma$) is
\begin{equation}
	\left|     
	\begin{matrix}	
		r_1  &  -r_1  &  r_3  & - r_3  \\
		- \tilde{\gamma} r_1 \! + \! Rm  &  \tilde{\gamma} r_1 \! + \! Rm  &  - \frac{q_z}{q_x} r_3 \! + \! Rm & \frac{q_z}{q_x} r_3 \! + \! Rm  \\
		\tilde{\gamma} e^{r_1}  &  \tilde{\gamma} e^{-r_1}  & \frac{q_z}{q_x} e^{r_3}  & \frac{q_z}{q_x} e^{-r_3}  \\
		(k_y \!\! + \! r_1 \! ) e^{r_1}  &  (k_y \!\! - \! r_1 \! ) e^{-r_1}  & (k_y \!\! + \! r_3 \! ) e^{r_3}  & (k_y \!\! - \! r_3 \! ) e^{-r_3}  
	\end{matrix}	
	\right| = 0 . \label{det}
\end{equation}
The second column is replaced by the average of the first and second, while the third is replaced by the average of the third and fourth
\onecolumngrid
\begin{equation}
	\left|     
	\begin{matrix}	
		r_1  &  0  &  r_3  & 0  \\
		- \tilde{\gamma} r_1 + Rm  &  Rm  &  - \frac{q_z}{q_x} r_3 + Rm & Rm  \\
		\tilde{\gamma} e^{r_1}  &  \tilde{\gamma} \cosh (r_1)  & \frac{q_z}{q_x} e^{r_3}  & \frac{q_z}{q_x} \cosh (r_3)  \\
		(k_y + r_1) e^{r_1}  &  k_y \cosh (r_1 ) + r_1 \sinh (r_1 ) & (k_y + r_3) e^{r_3}  & k_y \cosh (r_3 ) + r_3 \sinh (r_3 ) 
	\end{matrix}	
	\right| = 0 . \label{det2}
\end{equation}
	Adding $\tilde{\gamma}$ times the first line to the second, removing the second column to the fourth, and $r_3 /r_1$ times the first column to the third leads finally to
\begin{equation}
	\left|     
	\begin{matrix}	
		r_1  &  0  &  0  & 0  \\
		0 &  Rm  &  - \frac{q_z}{q_x} r_3 + \tilde{\gamma} r_3 & 0  \\
		\tilde{\gamma} \sinh (r_1 )  &  \tilde{\gamma} \cosh (r_1)  & \frac{q_z}{q_x} \sinh (r_3 ) - \tilde{\gamma} \frac{r_3 }{r_1 } \sinh (r_1 )  & \frac{q_z}{q_x} \cosh (r_3) - \tilde{\gamma} \cosh (r_1 )  \\
		r_1 \cosh (r_1 ) \! + \! k_y \sinh (r_1 )  &  r_1 \sinh (r_1 ) \! + \! k_y \cosh (r_1 ) & \left[ \begin{matrix} r_3 \cosh (r_3 ) + k_y \sinh (r_3 ) \\ - r_3 \cosh (r_1 ) - k_y \frac{r_3 }{r_1 } \sinh (r_1 ) \end{matrix}  \right] & \left[\begin{matrix} r_3 \sinh (r_3 ) + k_y \cosh (r_3 ) \\ - r_1 \sinh (r_1 ) - k_y \cosh (r_1 ) \end{matrix} \right] 
	\end{matrix}	
	\right| = 0 . \label{det3}
\end{equation}
Its determinant can be expressed analytically and leads to 
	\begin{equation}
		Rm = \frac{ \tilde{\gamma} \cosh (r_1 ) \left[ r_3 \sinh (r_3 ) + k_y \cosh (r_3 ) \right] - \frac{q_z}{q_x} \cosh (r_3 ) \left[ r_1 \sinh (r_1 ) + k_y \cosh (r_1 )  \right] }{\frac{ \tilde{\gamma} + \frac{q_z}{q_x}}{\tilde{\gamma} - \frac{q_z}{q_x}}  \left[ \cosh (r_3 ) \cosh (r_1 )  - 1 \right] + k_y  \left[ \frac{\sinh (r_3 )\cosh (r_1 )}{r_3} - \frac{\sinh (r_1 )\cosh (r_3 )}{r_1}\right] - \frac{ \frac{q_z r_1}{q_x r_3} + \tilde{\gamma} \frac{r_3 }{r_1 } }{\tilde{\gamma} - \frac{q_z}{q_x}} \sinh (r_3 ) \sinh (r_1 )}. \label{Rmga}
	\end{equation}
\twocolumngrid
For large values of $\gamma$, both values of $r_1$ and $r_3$ will be large too and $\cosh ( r_1 ) \simeq \sinh (r _1 ) \simeq 0.5 e^{r_1}$ (similarly for $r_3$) and the expression (\ref{Rmga}) can be approximated as 
\begin{equation}
	Rm \simeq \frac{\tilde{\gamma} r_3 }{1 - \frac{r_3}{r_1}}. \label{Rmgab}
\end{equation}
This expression may be re-arranged as
\begin{equation}
	Rm \simeq \frac{q_x }{ q_z } k \ g \left( \frac{\gamma}{k^2}, \eta _1 , q_x  \right), \label{Rmgab2}
\end{equation}
where the function $g$ is defined as
\begin{equation}
	g(x, \eta _1, q_x ) = x \frac{\sqrt{\frac{x+1+\eta _1}{1+ \eta _1 q_x^2}} }{ 1 - \sqrt{\frac{x+1+\eta _1}{(x+1) (1+ \eta _1 q_x^2)}}} . \label{funcg}
\end{equation}
From (\ref{Rmgab2}), the condition for optimal growth, $\partial \gamma / \partial k = 0$ leads to
\begin{equation}
	g - 2 x \frac{\partial g}{\partial x} = 0 , \label{optim}
\end{equation}
which implicitly provides $x$ as a function of $\eta _1$ and $q_x$. The condition that $Rm$ is positive is that $x > 1/q_x^2 -1 $, from equation (\ref{funcg}). For fixed values of $\eta _1$ and $q_x$, we have observed that there is only one solution $x_0$ to equation (\ref{optim}) in the range $]1/q_x^2 -1 ;  +\infty[ $.
Maintaining that value for $x_0$ implies that $\gamma \sim k^2$ and, from (\ref{Rmgab2}), that $Rm \sim k$. Stated otherwise, we have
\begin{align}
k & \sim Rm , \label{optk} \\
	\gamma &\sim  Rm^2 ,  \label{optRm}
\end{align}
modulo a function of $\eta _1$ (see appendix \ref{growth}) and $q_x$. 
This makes it a very fast dynamo: in dimensional terms the growth rate is proportional to $Rm$ times $U/H$. This is faster than the case of a uniform shear treated in \cite{RR84}, for which the growth rate is just independent of the electrical conductivity and proportional to $U /H$ (see their equation (11)) for a wavenumber proportional to the square-root of $Rm$ (see their equation (12)). To our knowledge, our dynamo is also the first example of a 'very fast dynamo', with a growth rate increasing (and unbounded) as the electrical conductivity is increased. 

\section{Growth rates of the uniform and localized shear flows}
\label{growth}

Let us now write the dynamo equations governing $P$ and $T$ for a general velocity profile $U(z)$. The idea is to understand why the dynamo obtained by Ruderman and Ruzmaikin \cite{RR84} is a fast dynamo (uniform shear) while our dynamo is very fast (localized Dirac shear). In \ref{derivationPT}, only two equations are changed, (\ref{cUBx}) and (\ref{ccUBz}) and become respectively
\begin{align}
	\left[\bnabla \times \left( {\bf u} \times {\bf B} \right) \right] _x &= k_x^2 U P' + k_x k_y U T + k^2 U' P, \label{cUBx2} \\
	\left[\bnabla \times \bnabla \times \left( {\bf u} \times {\bf B} \right) \right] _z &= - i k_x k^2 U  T -i k_y k^2 U' P . \label{ccUBz2}
\end{align}
The eigenvalue equations for $P$ and $T$, equations (\ref{eqP}) and (\ref{eqT}) are written here when $k_x =0 $ and $q_y = 0$, for a general velocity profile $U(z)$
\begin{align}
	\gamma P &= \left( 1 + \eta _1 q_x^2 \right) \left[ P'' - k^2 P \right] 
	+ i \eta _1 k_y q_x q_z T  , \label{eqP2} \\
	\gamma T &=  - i k_y U' P + T'' - k^2 \left( 1 + \eta _1 q_z^2 \right) T \nonumber \\
	& \hspace{20 mm}	+  i \eta _1 k_y q_x q_z \left[ P'' - k^2 P \right] . \label{eqT2}
\end{align}
In order to analyse the order of magnitude of $\gamma$, we shall consider both limits of small and large values of $\eta_1$. For small values of $\eta_1$ (which is the case considered in \cite{RR84}), 
at small wavenumber $k_y$, the dominant terms in (\ref{eqP2}) and (\ref{eqT2}) are
\begin{align}
	\gamma P &\simeq  
	 i \eta _1 k_y q_x q_z T  , \label{eqP3} \\
	\gamma T &\simeq - i k_y U' P  
	 . \label{eqT3}
\end{align}
Combining both equations and considering that $q_x$ and $q_z$ are of order unity leads to an estimate for the growth rate
\begin{equation}
	\gamma \sim \left( \eta _1 k_y^2 U' \right) ^{1/2} , \label{gamma_app}
\end{equation}
increasing with $k_y$. It is then limited by diffusion effects to an effective magnetic Reynolds number of order unity
\begin{equation}
	U' k_y^{-2} \sim 1. \label{ky_app}
\end{equation}
In the case of a linear profile \cite{RR84}, $U'$ is uniform and its value is equal to $Rm$, so that equations (\ref{gamma_app}) and (\ref{ky_app}) lead to
\begin{align}
	k_y & \sim Rm^{1/2} , \label{klinsmall} \\
	\gamma & \sim \eta _1^{1/2} Rm . \label{gammalinsmall} 
\end{align}
In the case of a Dirac function for $U'$, as considered in this paper, an estimate for $U'$ is directly related to the wavenumber, $U' \sim k_y Rm$. Hence equations (\ref{gamma_app}) and (\ref{ky_app}) now lead to
\begin{align}
        k_y & \sim Rm , \label{kDiracsmall} \\
	\gamma & \sim \eta _1^{1/2} Rm^2. \label{gammaDiracsmall}
\end{align}
Let us now consider the limit of large values for $\eta _1$. This limit has been found to possess a regular limit in the analytical solution developed in this paper. In this limit, it is useful to substitute $P'' - k^2 P$ in (\ref{eqT2}) using (\ref{eqP2}) to obtain
\begin{equation}
	\gamma \left[ T - i k_y \frac{\eta _1 q_x q_z}{1 + \eta _1 q_x^2} P \right] =  - i k_y U' P + T'' - k^2 \frac{ 1 + \eta _1 }{1 + \eta _1 q_x^2}  T .  \label{eqTcomb}
\end{equation}
In the limit of large $\eta _1$, equation (\ref{eqP2}) provides a relationship between $P$ and $T$ (on the right-hand side)
\begin{equation}
	P \sim k_y^{-1} T . \label{PT}
\end{equation}
For small values of $k_y$, equation (\ref{eqTcomb}) leads then to 
\begin{equation}
	\gamma \sim U' , \label{g_app_l}
\end{equation}
which is valid until diffusion effects dominate
\begin{equation}
	k_y^2 \sim U' . \label{k_app_l}
\end{equation}
For the linear profile $U' \sim Rm$, equations (\ref{g_app_l}) and (\ref{k_app_l}) lead to
\begin{align}
	k_y & \sim Rm^{1/2} , \label{klinlarge} \\
        \gamma & \sim Rm . \label{gammalinlarge}
\end{align}
For the Dirac function $U' \sim k_y Rm$, they lead to
\begin{align}
	k_y & \sim Rm , \label{kDiraclarge} \\
        \gamma & \sim Rm^2 . \label{gammaDiraclarge}
\end{align}
	In summary, the scaling (\ref{klinsmall}) and (\ref{gammalinsmall}) is that of \cite{RR84} for small $\eta _1$. For large $\eta _1$, we obtain here (\ref{klinlarge}) and (\ref{gammalinlarge}), which we have checked against the numerical solution of the eigenvalue problem with a linear velocity profile. Concerning the case of the Dirac velocity gradient profile studied in this paper, the scaling has been tested against the exact solution (\ref{Rmga}). In all cases, the growth rate is proportional to $\eta _1^{1/2}$ at small $\eta _1$ and then 
independent of $\eta _1$ at large values. In all cases, the uniform
gradient leads to a fast dynamo, while the Dirac gradient
leads to a very fast dynamo. This is due to the fact that the
velocity gradient can reach large values when large wavenumbers 
 are considered, which then implies that very fast dynamo action
is confined to a small region near a localized shear zone.

\end{document}